\documentclass[aps,pra,twocolumn]{revtex4-1}
\usepackage[utf8]{inputenc}
\usepackage[T1]{fontenc}
\usepackage{amsmath}
\usepackage{amssymb}
\usepackage{tikz}
\usepackage{bm}

\usetikzlibrary{calc}

\newcommand{\cuni}{\affiliation{Institute of Theoretical Physics, Faculty of Mathematics and Physics, Charles University, V~Hole\v{s}ovi\v{c}k\'{a}ch~2, Prague~8, 180~00, Czech Republic}}

\begin{document}

\title{Multi-photon above threshold ionization of multi-electron atoms and molecules using the R-matrix approach}

\author{Jakub Benda}
\email[]{jakub.benda@matfyz.cuni.cz}
\author{Zden\v{e}k Ma\v{s}\'{\i}n}

\cuni


\begin{abstract}
We formulate a computationally efficient time-independent method based on the multi-electron molecular R-matrix formalism. This method is used to calculate transition matrix elements for the multi-photon ionization of atoms and molecules under the influence of a perturbative field. The method relies on the partitioning of space which allows us to calculate the infinite-range free-free dipole integrals analytically in the outer region, beyond the range of the initial bound wave function. This approach is valid for an arbitrary order, that is, any number of photons absorbed both in the bound and the continuum part of the spectrum (below- and above-threshold ionization). We calculate generalized multi-photon cross sections and angular distributions of different systems (H, He, H$_{2}$, CO$_{2}$) and validate our approach by comparison with data from the literature.
\end{abstract}

\flushbottom
\maketitle

\thispagestyle{empty}

\section{Introduction}

Multi-photon ionization (MPI) and its variant resonance-enhanced multi-photon ionization (REMPI) in atoms and molecules have a range of important applications ranging from laser-induced plasma generation~\cite{sharma_2018,Vagin_2020}, chemical diagnostics~\cite{sharma_2020,boesl_1994,ryszka_2016}, chiral recognition~\cite{kastner_2017,beaulieu_2018,comby_2018,goetz_2019,lux_2012} to laser-filamentation~\cite{zvorykin,shutov_2017,dharmadhikari_2008,smetanin_2016,koga_2009}, harmonic~\cite{plummer_2002,Ackermann_2012,Ivanov_2008} and high-harmonic~\cite{brown_2012} generation and photoelectron spectroscopy~\cite{champenois_2016}. Given its practical importance accurate data on MPI are surprisingly scarce and striking discrepancies in the MPI cross sections remain in the literature despite recent advances in the experimental technology~\cite{sharma_2018}. This highlights the role of theory in supplying the missing data.

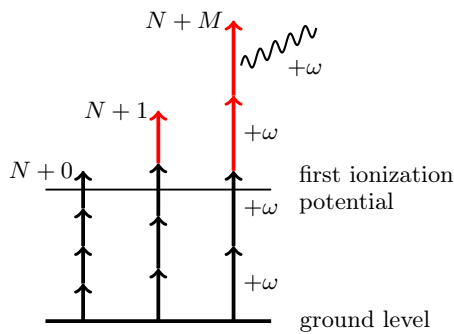
\begin{figure}[htbp]
    \centering
    \begin{tikzpicture}[scale=1.0]
        \draw[line width=0.50mm] (0,0) -- (3,0) ++(0.25,0) node[right] {ground level};
        \draw[line width=0.25mm] (0,1.75) -- (3,1.75) ++(0.25,0) node[right, align=left] {first ionization\\potential};
        \draw[->, line width=0.50mm] (0.5,0) -- (0.5,0.5);
        \draw[->, line width=0.50mm] (0.5,0.5) -- (0.5,1.0);
        \draw[->, line width=0.50mm] (0.5,1.0) -- (0.5,1.5);
        \draw[->, line width=0.50mm] (0.5,1.5) -- (0.5,2.0) node[left] {$N + 0$};
        \draw[->, line width=0.50mm] (1.5,0) -- (1.5,0.7);
        \draw[->, line width=0.50mm] (1.5,0.7) -- (1.5,1.4);
        \draw[->, line width=0.50mm] (1.5,1.4) -- (1.5,2.1);
        \draw[->, line width=0.50mm, red] (1.5,2.1) -- (1.5,2.8) node[left,black] {$N + 1$};
        \draw[->, line width=0.50mm] (2.5,0) -- node[right] {$+\omega$} (2.5,1.0);
        \draw[->, line width=0.50mm] (2.5,1.0) -- node[right] {$+\omega$} (2.5,2.0);
        \draw[->, line width=0.50mm, red] (2.5,2.0) -- node[right,black] {$+\omega$} (2.5,3.0);
        \draw[->, line width=0.50mm, red] (2.5,3.0) -- (2.5,4.0) node[left,black] {$N + M$};
        \draw[thick] plot[smooth,variable=\t,domain=0:360,samples=101] (
            {2.6 + \t/360},
            {3.4 + 0.5*\t/360 + 0.1*sin(5*\t)}
        ) ++(-0.15,-0.3) node[below] {$+\omega$};
    \end{tikzpicture}
    \caption{Illustration of the multi-photon transitions, including above-threshold ionization, calculable by the presented method. From left to right \(N+0\), \(N+1\) and \(N + M\) multi-photon ionization processes. Red arrows mark the \(M\) photons absorbed by the photo-electron ``in continuum''. In the special case of the \([N+M]\) REMPI scheme the first \(N\) photons excite the target to an intermediate bound state and the remaining \(M\) photon absorptions ionize the target without further photon absorptions in the continuum, thus corresponding to our \((N+M)+0\) case.
 }
    \label{fig:MPI}
\end{figure}

In the last two decades many experiments probing fundamental aspects of MPI have been carried out. Simulations of the multi-photon perturbative processes can provide valuable insight, whether due to the perturbative field used, or thanks to specific selection rules in effect that restrict the process to a specific N-photon transition. An example of the latter is the RABITT mechanism for measuring the photoionization time delays~\cite{Dahlstrom,Dahlstrom_JPB,vos_2018,SerovKheifets}, including its multi-sideband (multi-photon) variant~\cite{multirabitt}. Similarly, absorption of counter-rotating circularly polarised photons via several indistinguishable interfering pathways was shown to lead to characteristic electron vortices in the momentum angular distribution~\cite{vortices,vortices2,vorticesRMT,kerbstadt_2019}. Recently, a two-photon ATI of helium atoms by finite, few-femtosecond pulses was investigated using a time-dependent method~\cite{HePulses}, building on earlier monochromatic-pulse results from perturbation theory~\cite{Shakeshaft,Feng,Sanchez}. For a general introduction to the field of many-photon above-threshold ionisation (ATI) see~\cite{ATIreport} and references therein.

While MPI of atoms, particularly of the hydrogen-like type, has been thoroughly studied in the past~\cite{Manakov}, the literature on MPI of molecules is much more scarce due to the lack of the spherical symmetry and associated difficulties with the description of the intermediate and final wave functions of the system in the continuum. An exception to this rule is the multi-electron molecular R-matrix Floquet (RMF) approach~\cite{burke_molecular_2000} based on an earlier atomic RMF theory~\cite{Burke_1990} and applied to MPI of H$_{2}$~\cite{colgan_molecular_2001}. With the exception of two-photon cross-sections for molecular hydrogen~\cite{RitchieMcGuire,Morales09,Apalategui_2002,Demekhin} and nitrogen~\cite{LarsenN2} and general (\(N + 0\)) MPI of molecules in DFT approach~\cite{Toffoli2012}, multi-photon processes in molecules are studied using time-dependent approaches for dynamics in arbitrary external fields, see e.g.~\cite{vos_2018,SerovKheifets} and references in~\cite{MRMT}; this is computationally very demanding and doesn't provide direct access to the multi-photon transition matrix elements. The RMF approach for multi-electron systems~\cite{joachain_2007} is conceptually closest to the method developed here. Nevertheless, a general method specialized to the evaluation of accurate multi-photon matrix elements for molecules is not presently available.

In this work we bridge the gap in theoretical methodology by formulating, implementing and testing a time-independent ab initio R-matrix approach for the multi-photon ionization of multi-electron molecules. The stationary character of our method is crucial since it allows us to use high-level descriptions for the electronic structure of all bound and continuum wave functions involved while avoiding the computationally expensive time evolution of the wave function.

The one-photon molecular ionization problem has been formulated within the stationary R-matrix theory~\cite{Harvey} and implemented within the molecular package UKRmol+~\cite{UKRmolp}. It requires calculation of the final stationary photoionization state, which is the solution of the full Schrödinger equation with the incoming-wave boundary conditions. Two-photon ($2+0$) cross sections for photon energies below the single-photon ionization threshold have been calculated for molecular hydrogen~\cite{RMT,MRMT} by a similar approach.

In this work we generalize the molecular R-matrix photoionization methodology to all photon orders, see Fig.~\ref{fig:MPI}. The principal problem of the calculation is the evaluation of the free-free matrix elements. In the R-matrix formulation of photoionization this aspect is simplified by splitting of the coordinate space into the inner and outer region. In the inner region, exact exchange and multi-electron correlation effects are included using configuration interaction. In the outer region, where a single electron moves in the multipole potential of the residual molecule, the problem is treated using analytic techniques.

After a brief review of the computational method in the next two sections, we present results for multi-photon ionization of helium, molecular hydrogen and carbon dioxide across a continuous range of photon energies probing REMPI and non-resonant MPI. The full exposition of the theory is given in Appendix~\ref{sect:theory}. It is split into two parts. For clarity, in Sections~\ref{sect:bc} and~\ref{sect:free}, we first discuss the \((N + 1)\)-photon ionization, whose description requires fewer conceptual extensions of the one-photon ionization problem than the general \(N+M\) process. In Section~\ref{sect:ATI} we describe the general \((N + M)\)-photon process.

\section{Overview of the molecular R-matrix method}

A detailed description of the R-matrix theory and its molecular implementation has been given elsewhere~\cite{Burke,Harvey,UKRmolp,MRMT}. Here we limit ourselves to the definition of the key quantities needed for the development of the theory presented in this work.

The basic idea of this method is to divide the space by a sphere of radius $r=a$ (often called ``R-matrix radius'') into an inner region, where multi-electron interactions including exchange are important, and an outer region, where exchange and correlation between the continuum electron and the residual target are negligible. The R-matrix, constructed in the inner region, is the Green's function for the one-electron outer-region problem evaluated on the boundary and provides the link between both regions. In the following we denote all spin-space coordinates of the ($N+1$)/$N$ electrons by $\mathbf{X}_{N+1}/\mathbf{X}_{N}$.

In the outer region the \((N+1)\)-electron wave function can be written as a sum of direct products of the bound wave functions of the \(N\)-electron residual molecule and the one-electron wave functions of the continuum electron:
\begin{equation}
    \Psi(\mathbf{X}_{N+1}) = \sum_{p} \overline{\phi}_{p}^{\Gamma_{p}}(\mathbf{X}_{N};\hat{\mathbf{r}}_{N+1}\sigma_{N+1})\frac{1}{r}F_{p}(r),
    \label{eq:outerpsi}
\end{equation}
where $\overline{\phi}_{p}^{\Gamma_{p}}$ are the channel wave functions~\cite{MRMT} of space symmetry $\Gamma_{p}$ defined as the residual $N$-electron state coupled to the real spherical harmonic $X_{l_{p},m_{p}}(\hat{\mathbf{r}}_{N+1})$ and spin $\sigma_{N+1}$ of the continuum electron in the outer region; $F_{p}(r)$ is the radial channel wave function of the continuum electron.

In the inner region any solution $\Psi_{E}$ of the time-independent Schr\"{o}dinger equation for energy $E$ can be expressed as a linear combination of the R-matrix eigenstates $\psi_k$
\begin{eqnarray}
    \Psi_{E}(\mathbf{X}_{N+1}) = \sum_{k} A_{k}(E)\psi_k(\mathbf{X}_{N+1}),
    \label{eq:innerpsi}
\end{eqnarray}
where the form of the $A_{k}(E)$ coefficients depends on the choice of the asymptotic boundary conditions for the outer-region solution~\cite{Harvey}. The \((N+1)\)-electron eigenstates $\psi_k$ are expressed on the Close-Coupling level in terms of ``continuum configurations'' \(\mathcal{A}\Phi_i^N\eta_{ij}\) and ``\(L^2\) configurations'' \(\chi_m^{N+1}\) as
\begin{align}
    \psi_k(\mathbf{X}_{N+1}) = &\ \mathcal{A}\sum_{i,j}c_{ijk}\Phi_i^N(\mathbf{X}_{N})\eta_{ij}(\mathbf{r}_{N+1}, \sigma_{N+1}) \nonumber \\
    &+\sum_m b_{mk} \chi_m^{N+1}(\mathbf{X}_{N+1}),
    \label{eq:psiNp1}
\end{align}
where \(\mathcal{A}\) indicates the antisymmetrization operation, $\eta_{ij}(\mathbf{r}_{N+1}, \sigma_{N+1})$ are continuum spin-orbitals dependent on the position vector $\mathbf{r}_{N+1}$ and spin $\sigma_{N+1}$ with a non-zero amplitude on the R-matrix sphere, \(\chi_m^{N+1}\) are configurations not containing continuum orbitals. The summation over \(i\) runs over the subset of all residual ion eigenstates included in the model; \(j\) runs over those continuum orbitals that are coupled by symmetry to the respective residual ion states and \(m\) runs over configurations generated from the molecular orbitals fully contained inside the inner region. The coefficients \(c_{ijk}\) and \(b_{mk}\) are obtained by diagonalizing the Hamiltonian using equation
\begin{eqnarray}\label{eq:inner-ham}
    (\hat{H}+\hat{L})\psi_k = E_{k}\psi_k,
\end{eqnarray}
where $\hat{L}$ is the Bloch operator~\cite{Burke,UKRmolp}
\begin{equation}
    \hat{L} = \frac{1}{2} \sum_{i = 1}^{N + 1} \delta(r_i - a) \frac{\mathrm{d}}{\mathrm{d}r}.
    \label{eq:bloch}
\end{equation}

\section{Multi-photon ionization}

In this section atomic units are used exclusively. In the leading-order perturbation theory (LOPT) the total generalized \(K\)-photon ionization cross section for a fixed orientation of the molecule~\cite{Drake}
\begin{equation}
    \sigma_{fi}^{(K)} = w_{fi}^{(K)}/\phi^K = 2\pi (2\pi\alpha\omega)^K |M_{fi}^{(K)}|^2
\end{equation}
is proportional to the squared magnitude of the \(K\)-photon transition matrix element~\cite{Drake}
\begin{equation}
    M_{fi}^{(K)} = \langle \Psi_{f\bm{k}_f}^{(-)} | \hat{D}_{c_K} \hat{G}_{K-1}^{(+)} \dots \hat{D}_{c_2} \hat{G}_1^{(+)} \hat{D}_{c_1} | \Psi_i \rangle
    \label{eq:Mdef}
\end{equation}
and has dimension length\(^{2K}\) \(\times\) time\(^{K - 1}\).
Here \(\hat{D}_{c_k}\) is the projection of the dipole operator along the polarization \(c_k\) of the \(k\)-th absorbed photon, \(\Psi_i\) is the bound initial state of the molecule with energy \(E_i\), that is usually sufficiently well described by one of the inner-region R-matrix eigenstates, \(\Psi_{f\bm{k}_f}^{(-)}\) is the final stationary photoionization state as defined in~\cite{Harvey} and \(\hat{G}_k^{(+)}\) is the Green's operator of the full Hamiltonian \(\hat{H}\) of the system, evaluated at energy $E_{i}+k\omega$,
\begin{equation}
    \hat{G}_k^{(+)} = (E_i + k\omega - \hat{H} + \mathrm{i}0)^{-1} \,.
\end{equation}
For simplicity, we denote the aggregated energy of the first \(k\) absorbed photons as \(k\omega\), implying that all have the same energy, but in principle the energies of the photons can be different and the product \(k\omega\) replaced by a sum of their energies.

The formula~\eqref{eq:Mdef} for the transition matrix element is equivalent to solution of \(K - 1\) inhomogeneous Schr\"{o}dinger equations for stationary \emph{intermediate states} \(\Psi_{i + j\omega}^{(+)}\),
\begin{align*}
    (E_i + \omega - \hat{H}) \Psi_{i + \omega}^{(+)} &= \hat{D}_{c_1} \Psi_i \,, \\
    (E_i + 2\omega - \hat{H}) \Psi_{i + 2\omega}^{(+)} &= \hat{D}_{c_2} \Psi_{i+\omega}^{(+)} \,, \\
    \dots \\
    (E_i + (K-1)\omega - \hat{H}) \Psi_{i + (K-1)\omega}^{(+)} &= \hat{D}_{c_{K-1}} \Psi_{i+(K-2)\omega}^{(+)} \,,
\end{align*}
also discussed in~\cite{Toffoli2012}, followed by evaluation of the final dipole transition,
\begin{equation}
    M^{(K)} = \langle \Psi_{f\bm{k}_f}^{(-)} | \hat{D}_{c_K} | \Psi_{i+(K-1)\omega}^{(+)} \rangle \,.
    \label{eq:MKlast}
\end{equation}
The boundary conditions for the intermediate states are chosen to correspond to the physical outgoing-wave solution. As long as the combined photon energies are insufficient to ionize the target, the boundary condition is asymptotically zero. In the R-matrix formulation the bound intermediate states (excited states of the molecule) are accurately represented by the R-matrix eigenstates as long as the R-matrix radius is large enough to contain them fully: see~\cite{MRMT} where the $N+0$ cross sections for H$_2$ were calculated this way.

Once the combined energy of \(j\) photons exceeds the ionization threshold, the situation changes. The right-hand side that drives the equation for the first unbound intermediate state \(\Psi_{i+j\omega}^{(+)}\) is proportional to a bound state limited to the inner region, allowing us to use a purely outgoing solution in the outer region,
\begin{align}
    \Psi_{i + j\omega}^{(+)} &\rightarrow \frac{1}{r}\sum_p f_p(r) X_{l_p m_p}(\hat{\bm{r}}) \Phi_p  \,, \label{eq:Psijasy} \\
    f_p(r) &= a_p H_{l_p}^+(-Z/k_p, k_p r) \label{eq:fp} \,,
\end{align}
where \(Z\) is the residual ion charge, \(H_l^+(\eta,\rho)\) is the Coulomb-Hankel function, \(X_{lm}(\hat{\bm{r}})\) is the real spherical harmonic, \(\Phi_p\) is a state of the residual ion and the index \(p\) labels the asymptotic one-electron photoionization channels accessible after absorption of \(j\) photons. In writing~(\ref{eq:fp}) we have neglected higher multipoles of the electron-molecule interaction. However, the theory can be formulated to take them into account too, by means of an asymptotic expansion~\cite{Burke}. Nevertheless, in photoionization processes of neutral molecules the Coulomb interaction dominates and the higher multipoles can be safely neglected. It is very practical to consider closed channels too when the channel thresholds are approached from below, so that not only the continuum states but also highly excited bound states can ``leak'' from the inner region. For charged residual ions, the radial function of a closed channel is the exponentially decreasing real Whittaker function~\cite{DLMF},
\begin{equation}
    f_p(r) = a_p W_{Z/\kappa_p,l_p+1/2}(2 \kappa_p r) \label{eq:fpW} \,,
\end{equation}
where \(\kappa_p = \sqrt{-2 E_p}\) is the magnitude of the imaginary momentum associated with the closed channel \(p\).

The full exposition of the theory is given in Appendix~\ref{sect:theory}. Here we briefly note that the R-matrix method makes it possible to write equations for the intermediate states in the inner region that automatically contain the desired boundary condition in the outer region. The inner region solutions can be then unambiguously extended into the outer region by means of the special functions \(H_p^+\) and \(W_p\). The transition elements~\eqref{eq:MKlast} needed to calculate the cross sections are then calculated as a sum of contributions from the inner and outer region, employing the expansions~\eqref{eq:innerpsi} and~\eqref{eq:outerpsi}. In the outer region, this requires integration of highly oscillatory integrals from the boundary between the regions all the way to infinity. This integration is managed using asymptotic forms of the Coulomb (or Whittaker) functions and repeated integration by parts; see Appendix~\ref{sect:NDints} for details. When, for a given photon energy, the R-matrix radius chosen is insufficient for use of the asymptotic theory, numerical integration can be used for some radial interval as well.

\section{Results and discussion}

As the first application of the newly developed method we choose the hydrogen atom, where semi-analytic calculations have been performed by other authors long ago, see Fig.~\ref{fig:H-He-cross}a. We note that in this case all ``asymptotic'' forms discussed above and in Appendix~\ref{sect:theory} are actually exact solutions valid throughout all space which allows us to test our implementation (a custom one-electron code) and validate the presented R-matrix theory. After separation of the angular degrees of freedom, this problem is one-dimensional. The radial basis in the inner region consisted of 1000 equally spaced radial B-splines and extended up to radius \(a = 500\)~a.u. There are no surprises, the results obtained perfectly match the old calculations by Klarsfeld~\cite{Klarsfeld} and Karule~\cite{Karule}. In the calculations for H, He and H$_2$ (see below), we have not used the continuation of the bound intermediate states by means of equation~(\ref{eq:fpW}). Instead, a large enough R-matrix radius was used to contain these states.

\begin{figure*}[htbp]
    \centering
    \includegraphics[width=0.95\textwidth]{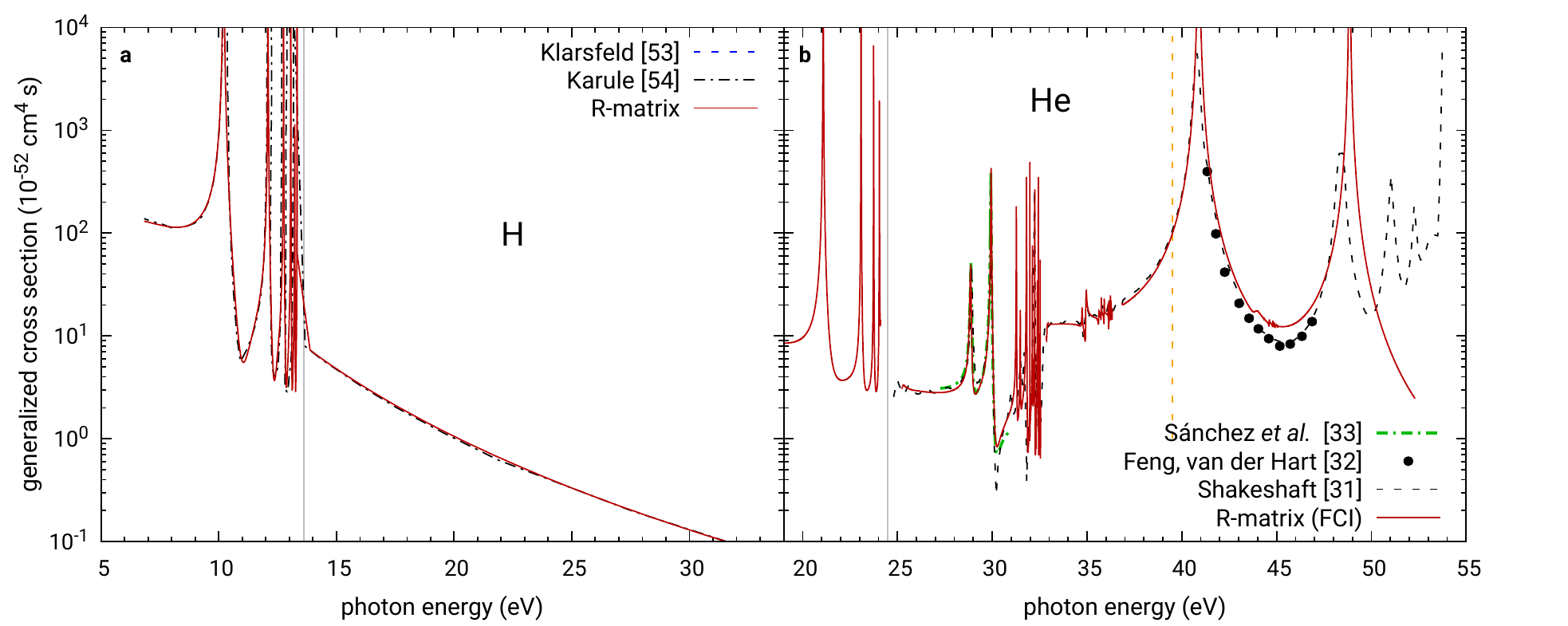}
    \caption{Left: Generalized cross section of two-photon ionization of the hydrogen atom. The vertical line marks the one-photon ionization threshold. Results are compared to earlier theoretical calculations of Klarsfeld~\cite{Klarsfeld} and Karule~\cite{Karule}. Right: Generalized cross section of two-photon ionization of the helium atom calculated in UKRmol+ with a Gaussian basis set used to represent the states of He$^+$. The solid grey vertical line marks the calculated one-photon single ionization threshold, while the broken yellow vertical line marks the two-photon double ionization threshold at~39.5~eV~\cite{Shakeshaft}. Results are compared to earlier theoretical calculations of Sánchez et al.~\cite{Sanchez}, of Feng and van der Hart~\cite{Feng} and of Shakeshaft~\cite{Shakeshaft}.}
    \label{fig:H-He-cross}
\end{figure*}

\subsection{Two-electron targets: Helium atom and H$_2$}

Still in the atomic domain, we performed a less trivial calculation of two-photon ionization of helium using UKRmol+~\cite{UKRmolp}. This required implementing the multi-channel version of our multi-photon approach into the UKRmol+ codes. The molecular R-matrix package was used to construct and diagonalize the inner-region Hamiltonian~\eqref{eq:inner-ham} for all required irreducible representations of the target's point group symmetry (here restricted to D$_{2h}$ as the largest available Abelian point group). Then, transition dipole elements between the R-matrix eigenstates are calculated for use in \(d_{\text{inn}}\), equation~\eqref{eq:dinnp}, and in the evaluation of the right-hand sides of the equations for the intermediate states, equation~\eqref{eq:masterEq}. Apart from this, the package also provides the transition dipole elements between the ionic states in \eqref{eq:Dqk}, the boundary amplitudes \(w_{kp}\), the real Gaunt's coefficients for equation~\eqref{eq:Qpws} and the necessary components of the final stationary photoionization state: the photoionization coefficients \(A_{pj}^{(-)}\) for equation~\eqref{eq:PsiAkminus} and the \(K\)-matrices needed for construction of the \(S\)-matrix in equation~\eqref{eq:Fqp}.

The size of the inner region was set to \(a = 100\)~a.u. As basic building blocks of the wave function of the residual ion we used Hartree-Fock orbitals of He$^+$ calculated in Psi4~\cite{PSI4} with the Gaussian basis set d-aug-cc-pVDZ. For the centre-of-mass-centred continuum basis we used the partial wave expansion up to \(\ell = 4\) and a radial basis set consisting of 200 evenly spaced B-splines. A full CI model was used for the $L^2$ expansion in equation~(\ref{eq:psiNp1}). To evaluate the outer region integrals we used 5 terms of the expansion~\eqref{eq:asyH} and \(P = 3\) in equation~\eqref{eq:Npp}. The results are shown in Fig.~\ref{fig:H-He-cross}b and compared with the available calculated results~\cite{Sanchez,Feng,Shakeshaft}. We can see a perfect agreement with earlier calculations below the first core excited resonance, i.e.\ resonant transition He\(^+\)(1s)--He\(^+\)(2s) in the residual ion, at around \(41\)~eV. The deviation occurring there and for higher photon energies is a direct consequence of the chosen Gaussian basis set, which is spatially limited and cannot represent diffuse excited states of the ion. Additionally, our calculations don't account for the double ionization channels opening at $39.5$~eV. These are expected limitations of the molecular package, which however do not contradict the validity of the presented approach.

The narrow intervals of energies around the one-electron ionization thresholds (in every channel) are the only problematic regions for this method. Below the threshold, one should see a series of Feshbach resonances converging to the threshold. This is difficult to represent accurately because the highly doubly-excited states responsible for these resonances are increasingly spatially extended requiring a very large R-matrix sphere to represent them sufficiently accurately. Additionally, the asymptotic expansion of the Coulomb-Hankel functions, written in terms of the argument \(\rho = kr\), becomes inapplicable close to the threshold (from above or from below) unless an extremely large R-matrix radius is used. Nevertheless, sufficiently far away from the threshold the theory works flawlessly and the results are very satisfying.

\begin{figure*}
    \centering
    \includegraphics[width=0.95\textwidth]{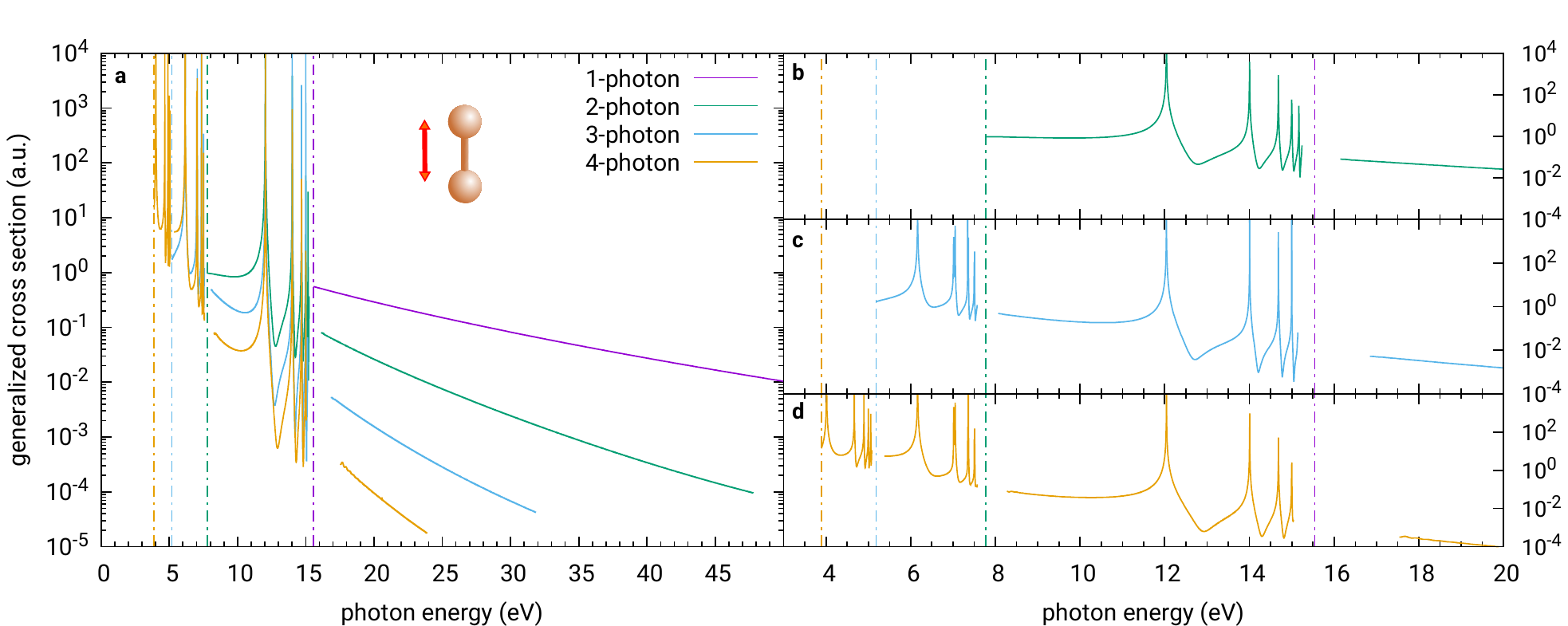}
    \caption{One-, two-, three- and four-photon generalized cross section for below- and above-threshold ionisation of the hydrogen molecule by a field polarised parallel with the molecular axis. The panels to the right provide details (from top to bottom) of the two-, three- and four-photon data around the multi-photon ionisation thresholds. The purple, green, blue and yellow vertical chain lines mark the one-, two-, three- and four-photon thresholds, respectively.}
    \label{fig:H2-multiphot}
\end{figure*}

In Fig.~\ref{fig:H2-multiphot} we compare one-, two-, three- and four-photon cross sections for ionisation of H$_2$ by a field polarised parallel to the molecular axis, with up to 3 photon absorptions in the continuum (high-energy section of the last curve). The repeated Feshbach resonance patterns below the thresholds of three-, two- and one-photon ionisation corresponding to excitation of a metastable state by several photons are well observable. There resonances correspond to intermediate excitations of the neutral molecule into one of the higher lying neutral bound states, ionized by the remaining photons. As before, in the vicinity of the thresholds the present method does not provide good results and the data were omitted from the plots. For this calculation we used static exchange model with the Hartree-Fock orbitals of the ion H$_2^+$ built from the atomic Gaussian basis set cc-pVDZ, R-matrix radius \(a = 150\)~a.u.\ and a continuum basis formed of 225 uniformly spaced radial B-splines and partial wave expansion up to \(\ell = 4\). The positions of the nuclei of both the initial neutral molecule and the residual ion were fixed at the equilibrium internuclear distance of H$_2$, which is 1.4 atomic units.

\begin{figure*}
    \centering
    \includegraphics[width=0.95\textwidth]{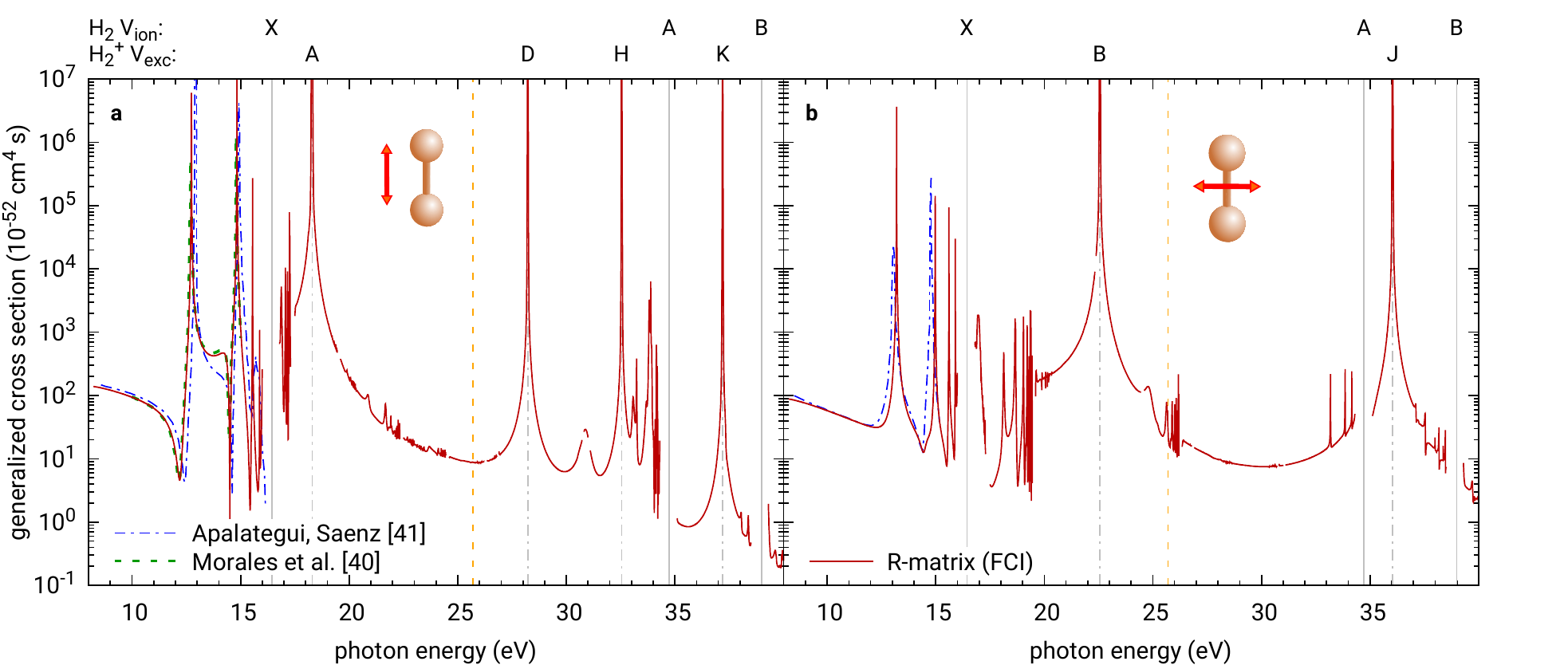}
    \caption{Generalized cross section of two-photon ionization of molecular hydrogen calculated using UKRmol+ with the atomic basis set aug-cc-pVTZ and full CI wave function model. Left panel: Field polarized parallel to the molecular axis. Right panel: Field polarized perpendicular to the molecular axis. Below-threshold results are compared to calculations of Apalategui and Saenz~\cite{Apalategui_2002} and Morales et al.~\cite{Morales09} The calculated one-photon ionization thresholds as well as the calculated positions of core-excited resonances allowed by symmetry are marked by grey vertical lines (solid and chain, respectively). The yellow broken line marks the calculated vertical non-sequential two-photon double ionization threshold at~25.7~eV.}
    \label{fig:H2-ATZ-FCI-oriented}
\end{figure*}

Further, in Fig.~\ref{fig:H2-ATZ-FCI-oriented} we present calculated cross sections for two-photon ionization of H$_2$ employing a full CI model with the aug-cc-pVTZ basis set that was used also in~\cite{MRMT} (called ``ATZ'' there). The polarization of the field was chosen parallel to the molecular axis (Fig.~\ref{fig:H2-ATZ-FCI-oriented}a) or perpendicular to it (Fig.~\ref{fig:H2-ATZ-FCI-oriented}b). The radius of the inner region was set to \(a = 100\)~a.u.\ in order to converge the results around the core-excited resonances well. To represent the continuum we included 150 evenly spaced B-splines of order 6 for partial waves up to \(\ell = 6\). The curve is interrupted at several energies around the one- and two-photon ionization thresholds (the latter are not marked), where the asymptotic expansion of Coulomb functions does not converge properly. The transition responsible for a core-excited resonance occurs within the residual ion rather than in the photo-electron, and so the momenta of the continuum functions in equation~\eqref{eq:doutp} are then very similar. This leads to problems with the asymptotic integrations, see in particular equation~\eqref{eq:I1explicit}, where the convergence of higher terms depends also on the assumption that \(|k - k'|r \gg 1\) in the denominator. However, extending the R-matrix radius and possibly reducing the number of terms in the asymptotic expansion in equation~\eqref{eq:Npp} (we used \(P=3\) in this calculation) mitigates this deficiency close to the resonances. Below the one-photon ionization threshold, the cross sections are compared to calculations of Apalategui and Saenz~\cite{Apalategui_2002} and Morales et al.~\cite{Morales09} with almost perfect agreement, disregarding small energy (horizontal) shifts arising from different relative energies of the ground state and the residual ion as obtained from the different methods. The only apparent disagreement is visible before the second Feshbach resonance in the first plot (parallel polarization), where the results of Apalategui and Saenz lack of the sharp ``shoulder'' below 15~eV visible in results of Morales et al.\ as well as in ours. This region was shown in~\cite{MRMT} to be sensitive to the molecular structure model used in the calculation. As in the case of helium, the description and energies of the higher excitations of H$_2^+$ manifesting as core-excited resonances are also most likely limited by the atomic basis set and the resonances might shift to lower energies when a more diffuse basis set is used. As the cross section is dominated by the contributions of the resonances (both core-excited and Feshbach ones), shifts in their positions translate to significant changes in the overall magnitude of the cross section.

\begin{figure*}
    \centering
    \includegraphics[width=0.95\textwidth]{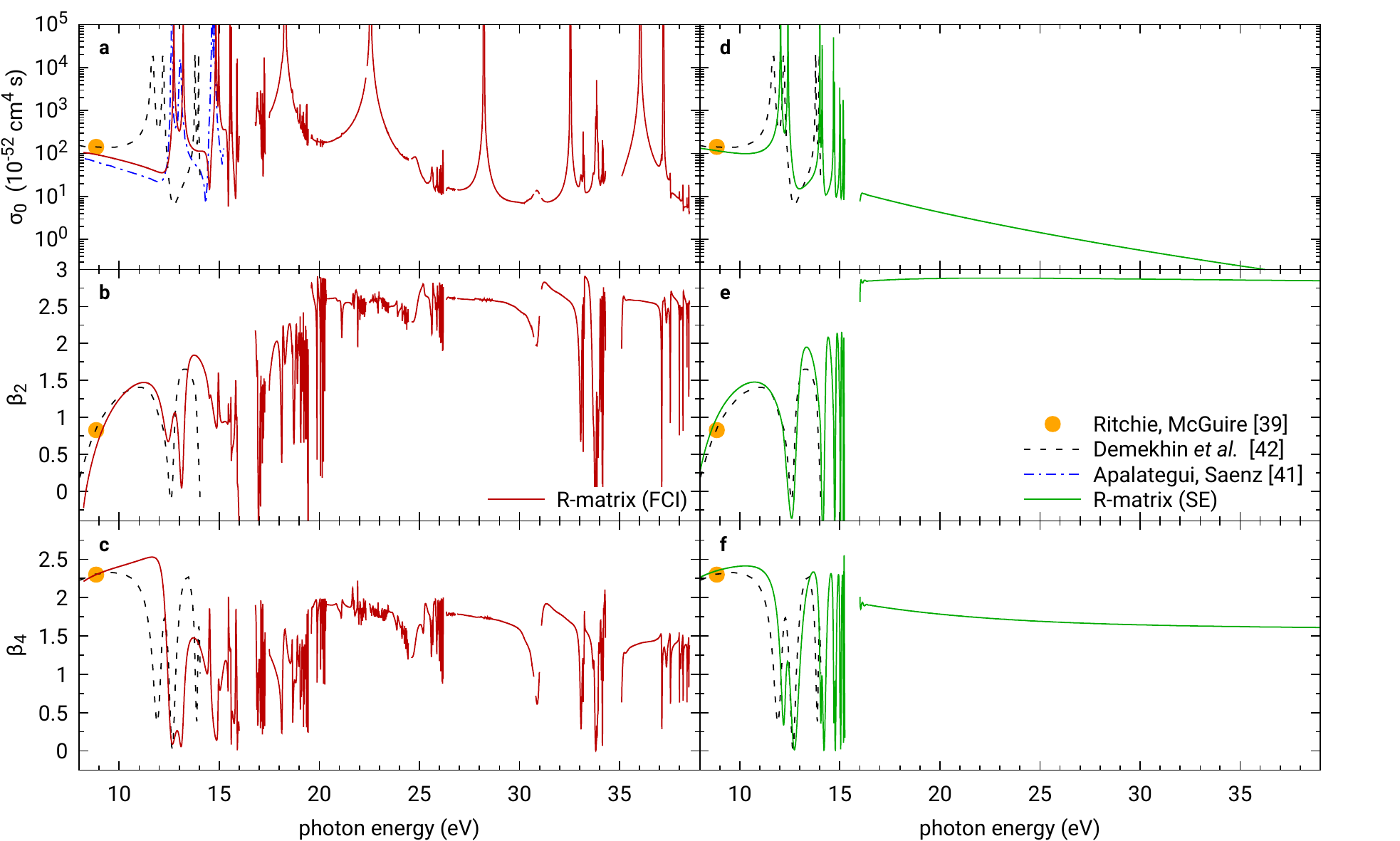}
    \caption{Laboratory-frame photoelectron angular distribution parameters for two-photon ionization of H$_2$. Left panels: With the same full CI model as in Fig.~\ref{fig:H2-ATZ-FCI-oriented}. Right panels: With the SE model as in Fig.~\ref{fig:H2-multiphot}. Top panels: Grand total cross sections averaged over all relative orientations of the molecule and the field polarization. Middle and bottom panels: Dimensionless asymmetry parameters for ionization into the ground state of H$_2^+$. The circle marks the theoretical result of Ritchie and McGuire~\cite{RitchieMcGuire}, chain curve the results of Apalategui and Saenz~\cite{Apalategui_2002} and the dashed curve the results of Demekhin et al.~\cite{Demekhin} The disagreement between our results and those of Apalategui and Saenz~\cite{Apalategui_2002}, possibly arising due to a typo in their codes, is discussed in Appendix~\ref{sect:oavg}.}
    \label{fig:H2-unoriented}
\end{figure*}

Figure~\ref{fig:H2-unoriented}a-c shows the molecular-orientation-averaged data: the total isotropic cross section \(\sigma_0\) and the asymmetry parameters \(\beta_2\) and \(\beta_4\), respectively, calculated by transforming the matrix elements~\eqref{eq:Mdef} to the spherical basis and performing averaging over molecular orientations parametrized by Euler angles as in~\cite{Toffoli2012, Demekhin}; see Appendix~\ref{sect:oavg}.
Alternatively, the isotropic cross section \(\sigma_0\) can be obtained directly in the real spherical harmonic basis and molecular frame by averaging only over polarizations \(\bm{\epsilon}\) of the photons and disregarding the dependence on the direction of the photoelectron momentum, by use of the formula~\cite{Zamastil,andrews_1977}
\begin{equation}
    \int \epsilon_{i} \epsilon_{j} \epsilon_{k} \epsilon_{l} \frac{\mathrm{d}\Omega}{4\pi}
    = \frac{1}{15} (\delta_{ij}\delta_{kl} + \delta_{ik}\delta_{jl} + \delta_{il}\delta_{jk}) \,.
    \label{eq:nnnn}
\end{equation}
The results below the one-photon ionization thresholds are compared to other published calculations~\cite{RitchieMcGuire, Apalategui_2002, Demekhin}. We see a good agreement with the calculation of Apalategui and Saenz~\cite{Apalategui_2002}; the small systematic difference in magnitude of the cross section before the first resonance is discussed in Appendix~\ref{sect:oavg}, where we show that it probably comes from a typo in the codes of Apalategui and Saenz evaluating their formula for the orientational average. We also observe a qualitative agreement with the calculations of Ritchie and McGuire~\cite{RitchieMcGuire} and Demekhin~et al.~\cite{Demekhin} Here, the quantitative differences also follow from very different ionization thresholds used; whereas Demekhin~et al.\ employ the experimental value of the \textit{adiabatic} first ionization threshold between ground vibronic states (15.43~eV), in the present calculation we use our calculated value for the \textit{vertical} threshold, which---for the employed model---comes out as 16.43~eV. The large 1~eV difference is responsible for most of the relative shift of the resonance features between the datasets. Furthermore, the quantitatively different behaviour is related to the choice of different quantum chemical models: The models used in~\cite{RitchieMcGuire} and~\cite{Demekhin} appear to provide results much closer to the static exchange model discussed earlier, see Fig.~\ref{fig:H2-unoriented}d-f, where also our calculated one-photon vertical first ionization threshold (15.55~eV) accidentally agrees better with the adiabatic experimental value. However, the more complete full CI model, albeit yielding a different vertical ionization potential, includes additional effects not captured by the previous simpler models.

\subsection{Multi-electron target: CO$_2$}

\begin{figure*}
    \centering
    \includegraphics[width=0.95\textwidth]{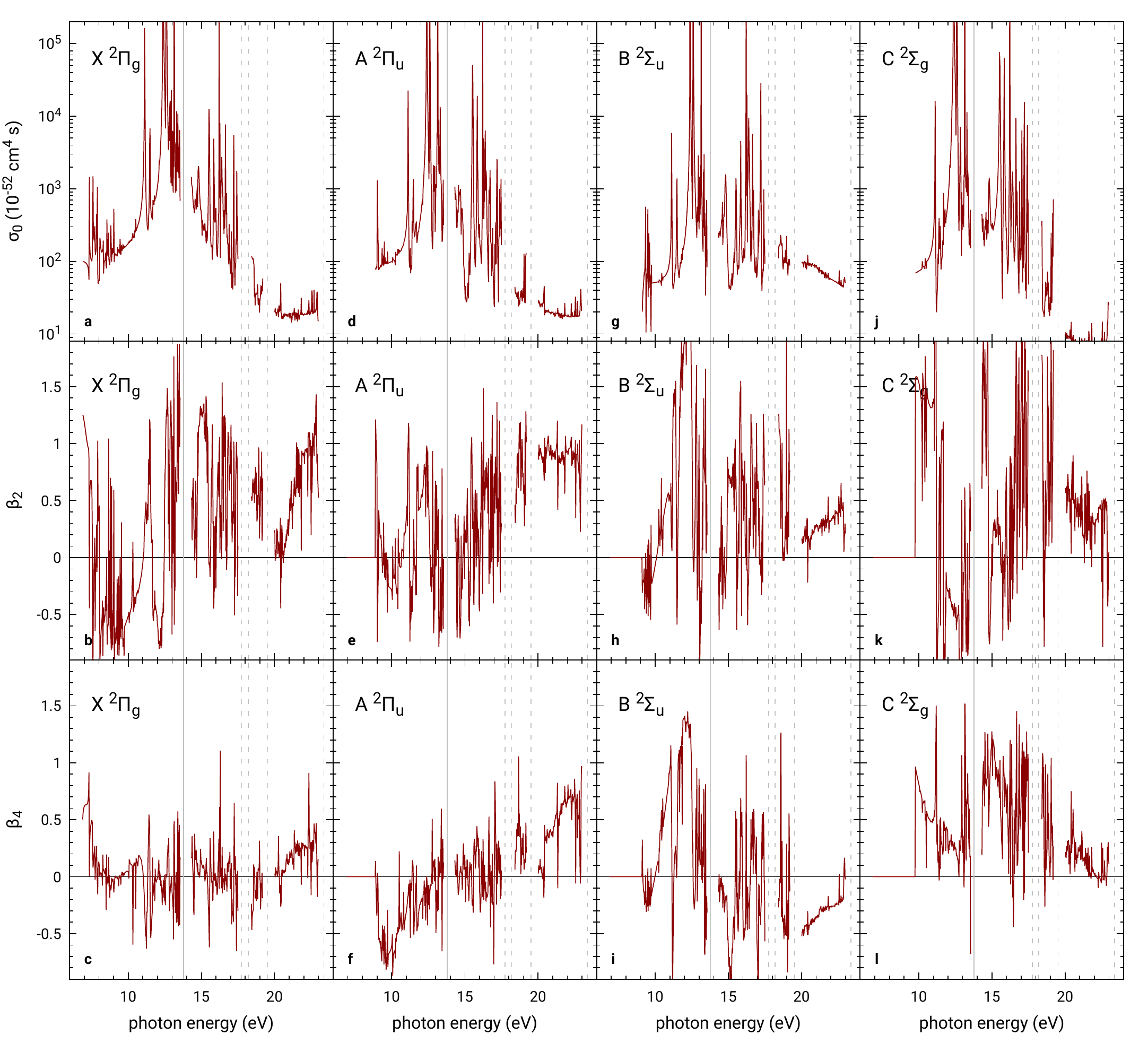}
    \caption{Laboratory-frame photoelectron angular distribution parameters for below- and above-threshold two-photon ionization of CO$_2$ into its first four ionic states. The solid grey vertical line marks the first one-photon ionization threshold at 13.78~eV, while the dashed lines mark further calculated one-photon thresholds for states A, B, C and D, indicating narrow energy windows with inaccurate results.}
    \label{fig:co2}
\end{figure*}

Finally, we calculate photoelectron angular distributions for the two-photon ionization of a comparatively larger molecule: CO$_2$, see Fig.~\ref{fig:co2}. Here we used a high-quality electronic model of the molecule introduced in~\cite{Harvey,co2hhg}, which was already demonstrated to yield very accurate one-photon cross section. The target model is based on the use of molecular orbitals of the neutral molecule in its equilibrium geometry (C--O distance 1.1621~\AA) optimized using the complete active space self-consistent field method (CASSCF) with cc-pVTZ basis. Here we focus on the low-energy region with photon energies up to $25$~eV (i.e. total absorbed energy up to $50$~eV) in which the continuum partial wave expansion converged for \(\ell = 6\). R-matrix radius of only \(a = 10\)~a.u.\ was sufficient and allowed to use Gaussian functions to represent the continuum orbitals. All molecular integrals were computed in quadruple precision using GBTOlib~\cite{UKRmolp} which allowed to avoid numerical linear dependencies and retain all continuum orbitals in the basis. Convergent single-photon cross sections were obtained including $300$ states of the residual ion in the Close-Coupling expansion~(\ref{eq:psiNp1}), see~\cite{co2hhg}. The radius of \(a = 10\)~a.u.\ is mostly too small for the asymptotic integrals to be applicable. Therefore the necessary one-dimensional outer-region integrations~\eqref{eq:integ2} have been performed numerically using the trapezoidal Romberg integration up to the radius \(b = 100\)~a.u. and analytically in the asymptotic part. Calculation with \(b = 200\)~a.u. was done for a subset of energies to check convergence away from the one-photon thresholds. To allow the intermediate bound states to extend beyond the inner region boundary we included closed channels~\eqref{eq:fpW} in the outer region wave function. As in the earlier one-photon calculation~\cite{co2hhg}, the energy of the ground state has been manually shifted by \(-1.1\)~eV to recover the experimental first ionization threshold 13.78~eV\cite{ChunCO2}.

\begin{figure}
    \centering
    \includegraphics[width=0.5\textwidth]{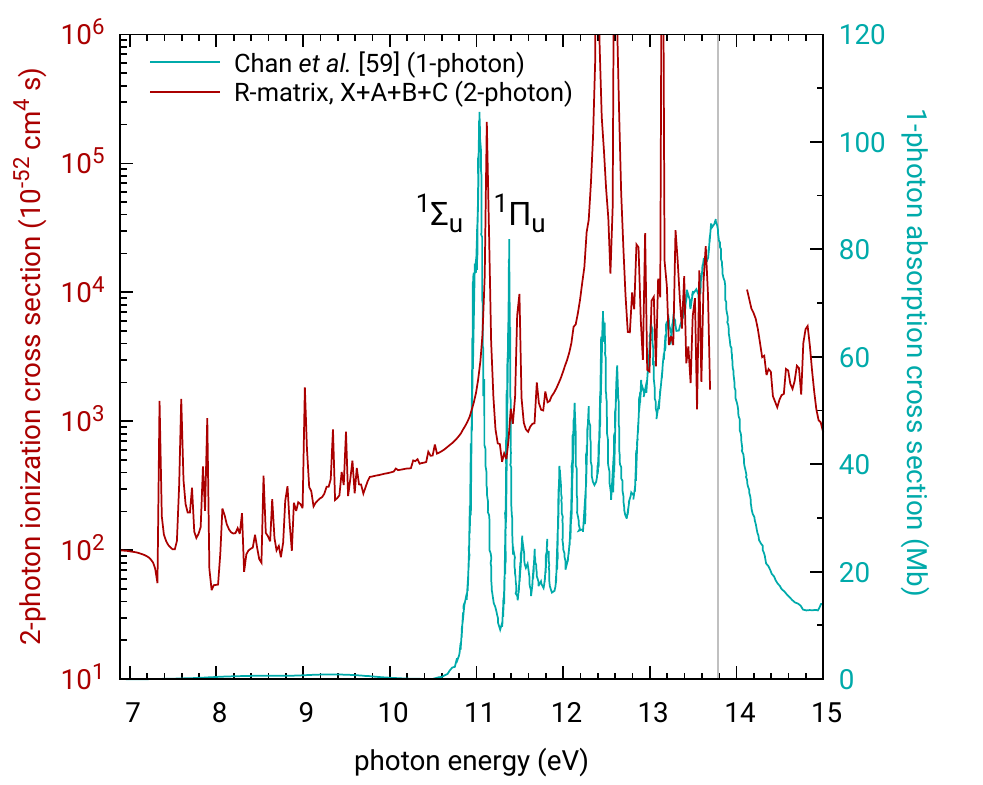}
    \caption{Summed two-photon isotropic partial ionization cross sections of CO$_2$ including the final states X, A, B, and C. The resonance structure in the below-threshold two-photon ionization is compared to below-threshold one-photon absorption measurement of Chan et al.~\cite{ChunCO2} The grey vertical line marks the one-photon ionization threshold. The first two resonances corresponding to dipole-allowed excitation of the neutral molecule to the singlet excited states \(^1\Sigma_u\) and \(^1\Pi_u\) are labeled in the plot.}
    \label{fig:co2-vs-absorb}
\end{figure}

In general, the partial isotropic cross sections for ionization into all four presented final states of CO$_2^+$ exhibit a rapid rise with the photon energy, reaching the maximum one or two eV before the above-threshold ionization threshold, from which they decrease. Qualitatively, this rise and decrease in the below-threshold two-photon ionization cross section is somewhat similar to the behaviour of the one-photon photoabsorption cross section measurement~\cite{ChunCO2} below the one-photon ionization threshold, see Fig.~\ref{fig:co2-vs-absorb}. The forest of resonances in the energy range 7--10 eV is associated with the two-photon thresholds of the residual ion states A, B, and C. The prominent pair of narrow isolated resonances that follows, located at approximately 11~eV, corresponds to intermediate excitation of the neutral molecule by the first photon to the lowest dipole-allowed states \({}^1\Sigma_u^+\) and \({}^1\Pi_u\) for parallel and perpendicular orientation of the molecule with respect to the polarization of the ionizing field, respectively. Right after these two resonances, further two-photon ionization channels open and the picture gets more complicated. The cross sections peak around the resonance associated with the excitation to the second \({}^1\Sigma_u^+\) state of neutral CO$_2$ at 12.4~eV. Beyond the first one-photon threshold the cross sections decrease and exhibit further resonance structures corresponding to higher autoionizing excited states, as well as resonances associated with further one-photon thresholds. In this calculation, 300 ionic states were included in the model, equation~\eqref{eq:psiNp1}, to correctly describe the polarization effects in the investigated range of energies. This is an order of magnitude more than the handful of states used for full CI calculations with He and H$_2$, with the first excited final state not even appearing before the ATI region. In the case of CO$_2$ these two-photon thresholds are much more densely spaced in energy and scattered all over the energy range of interest.

From the computational perspective, the most demanding part of the molecular multi-photon calculations is typically the evaluation of the molecular integrals needed in UKRmol+ to construct the molecular Hamiltonian. Among the discussed calculations, the molecular integrals were the most resource-demanding in the case of the full-CI model of H$_2$ with R-matrix radius of $100$~a.u., where their calculation took a little over 50 hours on a 40-core machine, producing 120 GiB of data. The remaining preparatory steps in UKRmol+ are typically faster; the diagonalization of the Hamiltonian and evaluation of transition dipole elements between its eigenstates for full-CI H$_2$ took 6 hours on the same machine. The evaluation of the two-photon ionization cross sections themselves for all 8000 distinct photon energies plotted in Fig.~\ref{fig:H2-unoriented}a-c and all $\sim$1700 final photoionization channels by the method proposed here amounted only to a little over 1 hour on a common 10-core workstation.

However, the complexity of the calculation of multi-photon ionization amplitudes discussed in this article is strongly affected by the number of channels, as well as by the above-threshold photon absorption order. For instance, disregarding the preparatory calculations in UKRmol+, the evaluation of the one- / two- / three- / four-photon ionization cross sections of H$_2$ in the static exchange model (with 12 outer region channels) for Fig.~\ref{fig:H2-multiphot} took 1~s / 1~s / 16~s / 1.5~h on a 10-core workstation, while the calculation of the two-photon ionization cross sections of CO$_2$ in Fig.~\ref{fig:co2} (up to $\sim$11000 channels, 680 distinct photon energies) took around 40 hours on a 36-core server machine, mostly because of the need for the numerical integration in the outer region due to the use of a very small R-matrix radius. For CO$_2$, the UKRmol+ stage was comparatively fast, consisting of 40~minutes for integral calculation in quadruple precision on 36~cores, less than 1~hour for Hamiltonian diagonalization and around 2 hours for calculation of the K-matrices needed in the outer region. Generally, it is advisable to extend the R-matrix radius as much as possible, so that the numerical integration in the outer region is avoided (or at least minimized) and a high resolution in energies becomes computationally cheap. This is feasible in UKRmol+ thanks to the availability of B-spline orbitals for construction of the continuum basis.

\section{Conclusion}

In this article we have presented a method for the calculation of multi-photon ionization amplitudes and cross sections for arbitrary photon orders. The method is based on the R-matrix approach, which accurately solves the time-independent Schrödinger equation in the close vicinity of the target system (atom, molecule, molecular cluster, \dots) and uses an analytic asymptotic expansion of the wave function elsewhere. The last point enables analytic treatment of the multi-dimensional oscillatory integrals that contribute to photon absorptions in above-threshold ionization. At the same time, the flexibility of the inner region allows us to treat complex targets using variational quantum-chemistry methods of the configuration-interaction type to describe the multi-electron dynamics.

Neglecting the channel coupling in the outer region, i.e.\ the long-range multipole potentials, is possible when the Coulomb interaction of the ejected electron with the residual ion is the dominant interaction. This narrows down the applicability of the method as presented here to situations where the ionized target has a non-zero charge or a negligible dipole moment; however, this is not a serious limitation, because most of the studied targets are neutral and the residual ions are singly charged. In any case the method can be straightforwardly extended to take into account channel coupling in the outer region by means of a multichannel asymptotic expansion of the outgoing solution~\cite{Burke}. The dipole integrals implied by this ansatz would have the same form as those worked out here.

The three cornerstones of the method are the following:
\begin{itemize}
    \item Embedding of the purely outgoing boundary condition in the Schrödinger equation via Bloch operator. (Appending~\ref{sect:theory})
    \item Reduction of the effective rank of the inhomogeneous system of equations by means of the Sherman-Morrison-Woodbury formula. (Appendix~\ref{sect:SMWformula})
    \item General algorithm for evaluation of the multi-dimensional asymptotic Coulomb-Green's integrals. (Appendix~\ref{sect:NDints})
\end{itemize}
For accurate results, this method presently requires the inner region to have a large enough size, so that the asymptotic formulas for Coulomb functions and their integrals can be applied. This can pose a problem close to thresholds, where linear momenta in some channels are small, the corresponding channel wave functions oscillate slowly and the asymptotic formulas require large distances to achieve the requested accuracy (or even validity). This can be tackled by further partitioning of the outer region into the ``transition'' part (say, for radii \(a\) to \(b\)) and the ``asymptotic'' part (from \(b\) to \(+\infty\)), where in the former one the integrals are evaluated numerically, while in the latter the analytic approach is used. In the presented results we have only used such splitting in the case of the two-photon ionization of CO$_2$ and aimed at large inner region radii otherwise.

We validated the theory by comparison of the generalized multi-photon cross sections to available published data for atomic hydrogen, helium and molecular hydrogen. We also provide original results for two-photon ionization of CO$_2$, below and above the ATI threshold. However, photoionization cross sections are not the only domain of applicability of this theory. The access to accurate transition amplitudes for polyatomic molecules can be used to study a range of interference phenomena (e.g.\ two- and multi-photon RABITT, electron vortices) using a time-independent approach, even though these have been traditionally the domain of time-dependent calculations~\cite{SerovKheifets,vos_2018,multirabitt}, or of approximative calculations that extend asymptotic wave functions all the way to the origin~\cite{Dahlstrom}. Our work opens the way to accurate calculations of multi-photon processes in multi-electron molecules. This is the direction of our further research.

\section*{Appendices}
\begin{appendix}

\section{Multi-photon ionization in R-matrix approach}
\label{sect:theory}

\subsection{Incorporation of the boundary condition}
\label{sect:bc}

In the following, the channel angular momentum numbers $l_{p}m_{p}$ will be collectively denoted by the index $p$ so the Coulomb-Hankel function and the real spherical harmonic become \(H_p^+(kr)\) and \(X_p\) respectively. We will not discuss the closed channels explicitly. The decreasing Whittaker function \(W_p\) is simply the Coulomb-Hankel function \(H_p^+\) with positive-imaginary momentum and has the same asymptotic expansion as \(H_p^+\). In the following text we do not distinguish these two functions and ``open channels'' can be understood to also include all weakly closed channels, whose penetration into the outer region needs to be considered to compensate for the finite inner region.

To embed the boundary condition~(\ref{eq:fp}) into the Schrödinger equation we proceed in the following way. First, we introduce the standard Bloch operator~\eqref{eq:bloch}
into the equation for the intermediate state by adding it to and subtracting from the Hamiltonian,
\begin{equation}
    \left[ E_i + j\omega - (\hat{H} + \hat{L}) + \hat{L} \right] \Psi_{i + j\omega}^{(+)} = \hat{D}_{c_j} \Psi_{i + (j-1)\omega}^{(+)} \,.
    \label{eq:schr1}
\end{equation}
We expand the wave function of the intermediate state in the R-matrix eigenstates \(\psi_m\) of the hermitian operator \(\hat{H} + \hat{L}\) in the inner region,
\begin{equation}
    \Psi_{i + j\omega}^{(+)} = \sum_{n} c_n \psi_n \,,
    \label{eq:Psijexp}
\end{equation}
and project the equation~\eqref{eq:schr1} on \(\psi_m\), resulting in
\begin{equation}
    \sum_n [(E_i + j\omega - E_m)\delta_{mn} + L_{mn}] c_n = \langle \psi_m | \hat{D}_{c_j} | \Psi_{i + (j-1)\omega}^{(+)} \rangle \,,
    \label{eq:schr2}
\end{equation}
where the matrix elements of the Bloch operator between the inner region eigenstates \(\psi_m\) and \(\psi_n\) can be expressed as
\begin{equation}
    L_{mn} = \frac{1}{2} \sum_p w_{pm}(a) w_{pn}'(a) \,,
\end{equation}
by means of the boundary amplitudes \(w_{pm}\) and radial derivatives \(w_{pn}'\) of those eigenstates when projected on the one-electron outer channels,
\begin{equation}
    w_{pm}(r) = \langle \tfrac{1}{r} \Phi_p X_p | \psi_m \rangle \,, \quad
    w_{pm}'(r) = \frac{\mathrm{d}}{\mathrm{d}r} \langle \tfrac{1}{r} \Phi_p X_p | \psi_m \rangle \,.
\end{equation}
Now, we can take advantage of the partially known asymptotic form~\eqref{eq:Psijasy} of the sought-after wave function. The boundary amplitude and the derivative of the radial channel wave function computed from the inner-region solution~(\ref{eq:Psijexp}) are
\begin{equation}
    f_p(a) = \sum_n w_{pn}(a) c_n \,, \quad
    f_p'(a) = \sum_n w_{pn}'(a) c_n \,.
    \label{eq:wdef}
\end{equation}
Comparing these expressions with~\eqref{eq:fp} and eliminating the unknown expansion coefficient \(a_p\) then yields
\begin{equation}
    \sum_n w_{pn}'(a) c_n = \sum_n \Lambda_{p}(a) w_{pn}(a) c_n \,,
    \label{eq:changeBC}
\end{equation}
where the boundary logarithmic derivative \(\Lambda_p(a)\) stands for
\begin{equation}
    \Lambda_p(a) = \frac{ \frac{\mathrm{d}}{\mathrm{dr}}H_{p}^+(k_{p}r)\vert_{r=a}}{ H_{p}^+(k_{p}a)} = \frac{k_p \frac{\mathrm{d}}{\mathrm{d\rho}}H_p^+(\rho)\vert_{k_{p}a}}{H_p^+(k_{p}a)} \,.
\end{equation}
Substituting~(\ref{eq:changeBC}) into the Schrödinger equation~\eqref{eq:schr2} leads to the final form
\begin{align}
    &\sum_n \left[(E_i + j\omega - E_m)\delta_{mn} + \frac{1}{2}\sum_p w_{pm}(a) \Lambda_p(a) w_{pn}(a)\right] c_n \nonumber \\
    &= \langle \psi_m | \hat{D}_{c_j} | \Psi_{i + (j-1)\omega}^{(+)} \rangle \,.
    \label{eq:masterEq}
\end{align}
This equation provides the correct inner region expansion coefficients of~\eqref{eq:Psijexp} compatible with the boundary condition~\eqref{eq:Psijasy}. The outer region expansion coefficients \(a_p\) can be evaluated from~\eqref{eq:wdef} and~\eqref{eq:fp} as
\begin{equation}
    a_p = \frac{\sum_n w_{pn}(a) c_n}{H_p^+(k_{p}a)} \,.
\end{equation}
The solution of equation~\eqref{eq:masterEq} can be optimized using the Sherman-Morrison-Woodbury formula. For details see Appendix~\ref{sect:SMWformula}.

\subsection{Free-free matrix elements}
\label{sect:free}

The final (detector) wave function can be expressed in a similar way as the intermediate states. In the inner region we can write the partial wave expansion of~\eqref{eq:innerpsi},
\begin{equation}
    \Psi_{f\bm{k}_f}^{(-)} = \sum_p \mathrm{i}^{l_p} \mathrm{e}^{-\mathrm{i}\sigma_p} X_{p}(\hat{\bm{k}}_f)
    \sum_k A_{pk}^{(-)}(k_f) \psi_k \,,
    \label{eq:PsiAkminus}
\end{equation}
where the partial-wave photoionization coefficients \(A_{pk}^{(-)}\) follow from the one-photon ionization theory~\cite{Harvey}, \(\sigma_p = \arg \Gamma(\ell_p + 1 - \mathrm{i}Z/k_f)\) is the Coulomb phase shift and \(\bm{k}_f\) the final momentum of the photoelectron after ionization into the residual state \(\Phi_f\). In the outer region it is
\begin{equation}
    \Psi_{f\bm{k}_f}^{(-)} = \frac{1}{r}\sum_p \mathrm{i}^{l_p} \mathrm{e}^{-\mathrm{i}\sigma_p} X_{p}(\hat{\bm{k}}_f)
    \sum_q F_{qp}^{(-)}(k_{q}r) X_{l_q m_q}(\hat{\bm{r}}) \Phi_q \,,
    \label{eq:Psifinn}
\end{equation}
where the radial function has the form~\cite{Burke}
\begin{equation}
    F_{qp}^{(-)}(kr) = \frac{-\mathrm{i}}{\sqrt{2\pi k_q}} \left(
        H_q^+(k_{q}r) \delta_{qp} - H_q^-(k_{q}r) S_{qp}^*
    \right)
    \label{eq:Fqp}
\end{equation}
and \(S_{qp}\) is the element of the S-matrix coupling the energy-accessible outer channels \(q\) and \(p\).
In the case of absorption of only one photon in the continuum (e.g. absorption of the second, IR photon in the RABITT scheme) the evaluation of the dipole matrix element~\eqref{eq:MKlast} can be split into inner region and outer region contributions,
\begin{equation}
    M = \sum_p \mathrm{i}^{-l_p} \mathrm{e}^{\mathrm{i}\sigma_p} X_{p}(\hat{\bm{k}}_f) (d_{\text{inn}, p} + d_{\text{out},p}) \,,
\end{equation}
\begin{align}
    d_{\text{inn}, p} &= \sum_{kj} A_{pk}^{(-)*}(k_f) c_j \langle \psi_k | \hat{D}_c | \psi_j \rangle \,, 
    \label{eq:dinnp} \\
    d_{\text{out}, p} &= \sum_{kq} \int\limits_a^{+\infty} F_{qp}^{(-)*}(k_{q}r) D_{cqk}(r) a_k H_k^+(k_{k}r) \mathrm{d}r \,,
    \label{eq:doutp}
\end{align}
The symbol \(D_{cqk}\) is the matrix element of the component \(c\) of the dipole operator between the states \(\Phi_q\) and \(\Phi_k\) of the residual ion and partial waves \(X_q\) and \(X_k\) of the ejected electron,
\begin{equation}
    D_{cqk}(r) = \langle \Phi_q X_q | \hat{D}_c | \Phi_k X_k \rangle = Q_{cqk}^{\text{ion}} + Q_{cqk}^{\text{pws}} r \,,
    \label{eq:Dqk}
\end{equation}
\begin{align}
    Q_{cqk}^{\text{ion}} &= \langle \Phi_q | \hat{D}_{c}^{\text{ion}} | \Phi_k \rangle \delta_{l_q l_k} \delta_{m_q m_k} \,, \\
    Q_{cqk}^{\text{pws}} &= \sqrt{\frac{4\pi}{3}} G_{1 m_c l_p m_p}^{l_q m_q} \langle \Phi_q | \Phi_k \rangle \label{eq:Qpws} \,.
\end{align}
The symbol \(G_{l_1 m_1 l_2 m_2}^{l_3 m_3}\) denotes the real Gaunt's coefficient (integral over three real spherical harmonics)~\cite{homeier_1996}. In equation~(\ref{eq:Dqk}) we have used the definition of the dipole matrix operator $\hat{D}_{c}$ and its separation into the ionic and one-electron parts:
\begin{equation}
    \hat{D}_{c} = \sqrt{\frac{4\pi}{3}}\sum_{i=1}^{N+1} r_{i}X_{1,c}(\bm{r}_{i}) = \hat{D}_{c}^{\text{ion}}+\sqrt{\frac{4\pi}{3}}r_{N+1}X_{1,c}(\bm{r}_{N+1}).
\end{equation}
To evaluate the contribution to the dipole matrix element from the outer region we need the integrals
\begin{equation}
    I^{(2)} = \int\limits_a^{+\infty} H_{l_1}^{s_1} r^m H_{l_2}^{s_2} \mathrm{d} r \,,
    \label{eq:integ2}
\end{equation}
where $s_{1}$ and $s_{2}$ stand for any combination of $+$ and $-$. These integrals can be evaluated for large values of \(a\) by replacing the Coulomb-Hankel functions by their asymptotic series, using the method of repeated integration by parts~\cite{Aymar,Cormier}. This method is explained in detail in Appendix~\ref{sect:NDints}.

\subsection{Higher-order above-threshold ionization}
\label{sect:ATI}

\begin{figure}[htbp]
    \centering
    \includegraphics[width=0.5\textwidth]{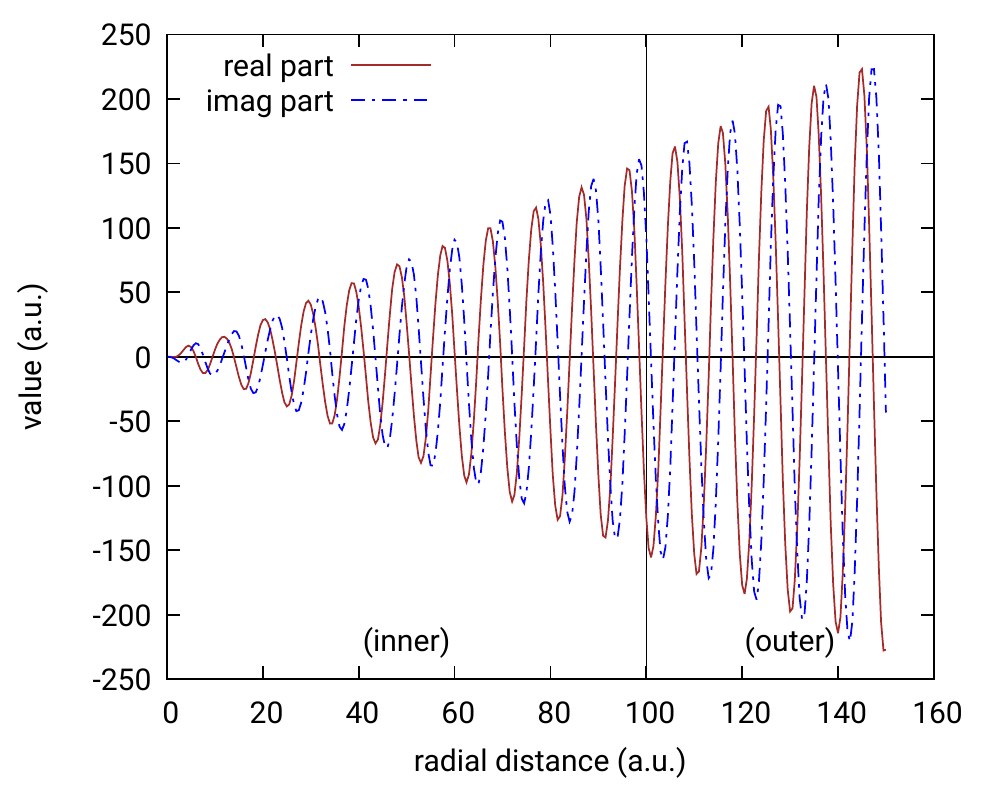}
    \caption{Real and imaginary part of the stationary intermediate-state wave function for above-threshold ionization of the hydrogen atom after absorption of one photon in the continuum for a specific energy. The vertical line as well as the labels ``(inner)'' and ``(outer)'' mark the division into R-matrix regions.}
    \label{fig:H-intermediate}
\end{figure}

When we go to higher orders (or energies) and it becomes possible that one or more photons leading to a given \textit{intermediate} state are absorbed in the continuum, the right-hand side of equation~(\ref{eq:schr1}) is non-zero in the outer region or even polynomially diverges with radius. This behaviour is transferred also to the solution itself, see Fig.~\ref{fig:H-intermediate}. The method then needs a generalization. Now, there is a source term both in the inner region and in the outer region. The value of the wave function at the R-matrix region boundary is a combination of these two sources. The inner region source, as before, generates an outgoing Coulomb-Hankel wave in each accessible channel. In addition, given that the outer region channels are not coupled, the contribution from the outer region to channel \(p\) can be directly obtained employing the Coulomb-Green's function
\begin{equation}
    g_p^{(+)}(r, r') = -\frac{2}{k_p} F_p(k_{p}r_<) H_p^+(k_{p}r_>) \,,
    \label{eq:gp}
\end{equation}
where $F_{p}(k_{p}r)$ is the regular Coulomb function~\cite{DLMF}. The equation~\eqref{eq:gp} is only valid for open channels. For energy-forbidden channels \(p\) one should use the negative-energy Green's function involving the real Whittaker functions~\cite{cgreenf}. In this work, though, we disregard contribution of closed channels to the 3-photon (and higher) ionization and limit ourselves to large R-matrix radii instead in such cases. For this reason we do not use the negative-energy Green's function, even though the extension is straightforward.

Combining the two contributions to the outer wave function gives
\begin{equation}
    f_p(r) = a_p H_p^+(k_{p}r) + \int\limits_a^{+\infty} g_p^{(+)} (r, r') \langle \Phi_p X_p | \hat{D}_{c_j} | \Psi_{i + (j-1)\omega}^{(+)} \rangle \mathrm{d}r' \,.
    \label{eq:fpATI}
\end{equation}
At the inner region boundary we then have
\begin{align}
    f_p(a) &= a_p H_p^+(k_{p}a) + \beta_p F_p(k_{p}a) \,, \label{eq:fpa} \\
    f_p'(a) &= a_p k_p \frac{\mathrm{d}}{\mathrm{d\rho}}H_{p}^{+}(\rho)\big\vert_{k_{p}a} + \beta_p k_p \frac{\mathrm{d}}{\mathrm{d\rho}}F_{p}(\rho)\big\vert_{k_{p}a} \,,
\end{align}
where
\begin{equation}
    \beta_p = -\frac{2}{k_p} \int\limits_a^{+\infty} H_p^+(k_{p}r) \langle \Phi_p X_p | \hat{D}_{c_j} | \Psi_{i+(j-1)\omega}^{(+)} \rangle \mathrm{d}r' \,.
    \label{eq:beta}
\end{equation}
Evaluation of this integral is discussed in Appendix~\ref{sect:NDints}. At the same time, the channel function \(f_p\) and its derivative need to satisfy equation~\eqref{eq:wdef}, leading to a generalization of~\eqref{eq:changeBC},
\begin{align}
    \sum_n &w_{pn}'(a) c_n = \sum_n \Lambda_{p}(a) w_{pn}(a) c_n \nonumber \\
        &+ \beta_p (k_p \frac{\mathrm{d}}{\mathrm{d\rho}}F_{p}(\rho)\vert_{k_{p}a} - \Lambda_p(a) F_p(k_{p}a)) \,.
\end{align}
The additional term that reflects the effect of the outer region source in the inner region does not depend on the sought expansion coefficients \(c_n\) and hence can be transferred to the right-hand side, yielding the final, fundamental equation of this method
\begin{align}
    &\sum_n \left[(E_i + j\omega - E_m)\delta_{mn} + \frac{1}{2}\sum_p w_{pm}(a) \Lambda_p(a) w_{pn}(a)\right] c_n \nonumber \\
    &= \langle \psi_m | \hat{D}_{c_j} | \Psi_{i + (j-1)\omega}^{(+)} \rangle \nonumber \\
    &- \frac{1}{2}\sum_p w_{pm}(a) \beta_p (k_p \frac{\mathrm{d}}{\mathrm{d\rho}}F_{p}(\rho)\vert_{k_{p}a} - \Lambda_p(a) F_p(k_{p}a))\,.
    \label{eq:masterEqATI}
\end{align}
The radial integration in \(\langle \psi_m | \hat{D}_j | \Psi_{i + (j-1)\omega}^{(+)} \rangle\) is limited to the inner region since the state \(\psi_m\) is restricted to it. The additional term on the right-hand side compensates for this truncation. Given the updated boundary formula~\eqref{eq:fpa}, the formula for the outer region expansion coefficients has to be generalized to
\begin{equation}
    a_p = \frac{\sum_n w_{pn}(a) c_n - \beta_p F_p(k_{p}a)}{H_p^+(k_{p}a)} \,.
\end{equation}
Together with~\eqref{eq:fpATI}, we now have the complete prescription for the intermediate state (given by \(c_n\) and \(a_p\)) in the whole configuration space that can be used in the evaluation of the final transition dipole elements, or as a seed for intermediate states of even higher order. Of course, when the right-hand side of~\eqref{eq:schr1} \emph{is} limited to the inner region, then \(\beta_p = 0\) and the equations in~\ref{sect:ATI} simplify to their restricted forms introduced earlier in \ref{sect:bc}.

Evaluation of \(d_{\text{out},p}\) (as well as of \(\beta_p\), which has the same form) gets progressively more and more complicated with the increasing order, requiring integrals of dimension increasing with the number of photons absorbed in the continuum, e.g.\ for 3 and 4 photons absorbed above the one-photon ionization threshold:
\begin{align}
    I^{(3)} &= \int\limits_a^{+\infty}\int\limits_a^{+\infty}
    H_{l_1}^{s_1} r_1^{m_1} g_{l_2}^{(+)} r_2^{m_2} H_{l_3}^{s_2}
    \mathrm{d} r_1 \mathrm{d} r_2 \,, \\
    I^{(4)} &= \int\limits_a^{+\infty} \int\limits_a^{+\infty} \int\limits_a^{+\infty}
    H_{l_1}^{s_1} r_1^{m_1} g_{l_2}^{(+)} r_2^{m_2} g_{l_3}^{(+)} r_3^{m_3} H_{l_4}^{s_4}
    \mathrm{d} r_1 \mathrm{d} r_2 \mathrm{d} r_3 \,,
\end{align}
etc. For simplicity, the radial and momentum-dependent arguments have been suppressed, but it should be clear that the Green's function separates parts of the expression that depend on adjacent \(r_i\) and \(r_{i+1}\). These nested integrals arise because the high-order intermediate states entering the formulas~\eqref{eq:MKlast} and~\eqref{eq:beta} no longer have the simple asymptotic proportional to \(a_p H_p^+\) and contain, as the above-threshold absorption order increases, an increasing number of integrals over the Green's function, see equation~\eqref{eq:fpATI}.
Evaluation of the required multi-dimensional integrals is discussed in Appendix~\ref{sect:NDints}.

The overall asymptotic complexity of the method is exponential, \(O(\exp N)\), in the number of absorptions below the ionisation threshold due to the exponentially rising number of possible combinations of absorbed photon polarizations, and even factorial, \(O(M!)\), in the number of absorptions in the continuum, above the ionisation threshold, as discussed in Appendix~\ref{sect:NDints}. However, as the practical applications of perturbative theory are generally limited to low orders anyway, the rapidly rising complexity for high-order above-threshold ionisation is not particularly worrying.

\section{Efficient solution of ``diagonal plus low rank'' equation}
\label{sect:SMWformula}

In the standard R-matrix theory of collisions the inversion of the inner-region Hamiltonian, equation~\eqref{eq:inner-ham}, needs to be performed only once. However, since equation~\eqref{eq:masterEq} needs to be solved for each photon energy independently, it pays off to implement this solution efficiently. As written, the matrix of the set of equations has rank equal to the number of inner region eigenstates, which may be a very large number \(N\). Solution of such equation has the asymptotic complexity \(O(N^3)\). Fortunately, it has a very specific form, sometimes called ``diagonal plus low rank'' matrix, where the low rank refers to the second term of the matrix in equation~\eqref{eq:masterEq} being constructed from boundary amplitude matrices, which have one of the dimensions equal to the number of photoionization channels \(n\), while typically \(n \ll N\). In this case, one can make use of the Sherman-Morrison-Woodbury formula~\cite{Higham},
\begin{equation}
    (\mathbf{A} + \mathbf{U} \mathbf{C} \mathbf{V})^{-1}
    = \mathbf{A}^{-1} - \mathbf{A}^{-1} \mathbf{U} (\mathbf{C}^{-1} + \mathbf{V} \mathbf{A}^{-1} \mathbf{U})^{-1} \mathbf{V} \mathbf{A}^{-1} \,,
\end{equation}
which allows expressing the solution of the \(N\)-by-\(N\) system using the solution of an \(n\)-by-\(n\) one, where \(N\) is the number of inner region eigenstates and \(n\) is the number of the outer region channels. In this case the solution is
\begin{equation}
    \mathbf{c} = \left(\mathbf{P} - \mathbf{P}
    \mathbf{w}^\top
    \left[
    \frac{1}{\frac{1}{2}\bm{\Lambda}} + \mathbf{w} \mathbf{P} \mathbf{w}^\top
    \right]^{-1}
    \mathbf{w} \mathbf{P}
    \right) \mathbf{D} \,,
    \label{eq:SMW-formula-RM}
\end{equation}
where \(\mathbf{P} = (E_{\text{tot}} - \mathbf{E})^{-1}\) is a diagonal \(N\)-by-\(N\) matrix. Note that the rank of the inner inversion in~\eqref{eq:SMW-formula-RM} needs to be limited to the number of open channels \(n_o \le n\). Now the most time-consuming operation becomes the construction of the matrix \(\mathbf{w} \mathbf{P} \mathbf{w}^\top\), consisting of \(O(n_o^2 N)\) multiplications, which for $n_{o} < N$ outperforms the \(O(N^3)\) operations required to solve the original system, particularly for a small number of open channels.

\section{Evaluation of multi-dimensional Coulomb-Green dipole integrals}
\label{sect:NDints}

Multi-dimensional integrals of products of coordinate powers, Coulomb-Hankel functions and Coulomb-Green's functions from a large radius \(a\) to infinity in each coordinate can be expressed as a sum of ``triangular'' integrals over \(a < r_1 < r_2 < \dots < r_M < +\infty\), in order to fix the assignment of the coordinates in the Green's function~\eqref{eq:gp}. Then, the regular Coulomb functions \(F_l\) are rewritten using Coulomb-Hankel functions \(H_l^\pm\):
\begin{equation}
    F_{l} = \frac{1}{2\mathrm{i}}\left[H_{l}^{+} - H_{l}^{-} \right].
\end{equation}
All Coulomb-Hankel functions are in turn expanded using the asymptotic series~\cite{DLMF}
\begin{equation}
    H_l^{s}(\eta, \rho) \sim \mathrm{e}^{\mathrm{i} s\phi_l(\eta,\rho)}
    \sum_{n = 0}^\infty \frac{(a)_n (b)_n}{n! (2\mathrm{i}s\rho)^n} \,,
    \label{eq:asyH}
\end{equation}
where \(a = 1 + l + \mathrm{i} s\eta\), \(b = -l + \mathrm{i} s\eta\), \(\eta = -Z / k\), \(\rho = kr\), \(\phi_l(\eta,\rho) = \rho - \eta\log(2\rho) - \pi l / 2 + \sigma\), \(\sigma=\arg\Gamma(l+1+\mathrm{i}\eta)\) and \(s\) is either \(+1\) or \(-1\).
As a result, the final integrand in~\eqref{eq:beta} will become a product of oscillatory exponentials and (possibly negative) coordinate powers. In general, such integral does not have a well defined value. We circumvent this by introduction of an auxiliary damping factor \(\exp(-cr_i)\) with positive constant \(c\) into the integrand for each coordinate \(r_i\) as suggested in~\cite{Starace}. Following the integration we perform the limit \(c \rightarrow 0+\); the value of the integral stays finite and well-defined. All formulas in this section are to be understood in this limit.
The prototypical one-dimensional integral
\begin{equation}
    I[f, \theta]_a \equiv \int\limits_a^{+\infty} f(r) \mathrm{e}^{\mathrm{i}\theta(r)} \mathrm{d}r \,,
    \label{eq:I1D}
\end{equation}
with \(\theta(r) = s_1\theta_1(r) + s_2\theta_2(r)\), where \(\theta_i(r) = k_i r + Z\log(2k_ir)/k_i - \pi l_i / 2 + \sigma_i\),
can be evaluated by repeated integration by parts (limited to \(P\) iterations) as done in~\cite{Aymar,Cormier}, leading to
\begin{equation}
    I[f, \theta]_a = \mathrm{e}^{\mathrm{i}\theta(a)} \gamma(a) \sum_{t = 0}^{P} T_{t}(a) \,,
    \label{eq:Npp}
\end{equation}
where \(\gamma(r) = \mathrm{i}/\theta'(r)\) and the terms \(T_t(r)\) satisfy the recursive definition
\begin{align}
    T_0(r) &= f(r) \,, \\
    T_{t + 1}(r) &= \frac{\mathrm{d}}{\mathrm{d}r} \left( \gamma(r) T_t(r) \right) \,.
\end{align}
The primes indicate derivatives with respect to \(r\). For evaluation of the terms \(T_t\),
the derivatives of \(f(r)\) at \(r = a\) need to be known, which is easy for the one-dimensional case~\eqref{eq:I1D},
where \(f(r)\) is simply a (small, or even negative) power of \(r\). The sum in~\eqref{eq:Npp} is asymptotic series similar to~\eqref{eq:asyH}; higher terms are proportional to high derivatives of negative powers of the coordinate and to high derivatives of the phase \(\theta(r)\) at large distances, which rapidly decrease in magnitude with each new term, provided that the evaluation radius \(r = a\) is sufficiently large. For modest radii, though, considering fewer terms leads to better results.

Nested ``triangular'' integrals of the type
\begin{equation}
    \int\limits_a^{+\infty}
    \dots
    \int\limits_{r_3}^{+\infty}
    f_2(r)
    \mathrm{e}^{\mathrm{i}\theta_2(r)}
    \int\limits_{r_2}^{+\infty}
    f_1(r)
    \mathrm{e}^{\mathrm{i}\theta_1(r)}
    \mathrm{d} r_1
    \mathrm{d} r_2
    \dots
    \mathrm{d} r_N
\end{equation}
arising in absorption of at least two photons in the continuum can be calculated similarly by recursion, based on the observation that
\begin{equation}
    I[f_1, \theta_1]_r = g_1(r) \mathrm{e}^{\mathrm{i}\theta_1(r)}  \,,
\end{equation}
and so for a two-dimensional integral
\begin{equation}
    I[f_2 I[f_1, \theta_1], \theta_2] = I[f_2 g_1, \theta_1 + \theta_2] \,.
\end{equation}
That is, the integrand at any level has the same form ``function times exponential'' and the same approach can be used to evaluate the integral as for lower orders. For the second-order integral, the only additional information needed to evaluate it is the knowledge of the value and derivatives of \(\tilde{f}_2 = f_2 g_1\) at \(r = a\), which can be precalculated before evaluating the second integral. The need of \(P\) derivatives of \(\tilde{f}_2\) translates, via the chain rule, to the need to know \(P\) derivatives of \(g_1\), which in turn just raises the number of derivatives of \(f_1\) that need to be precalculated beforehand, because derivatives of the terms \(T_t\) are then needed, too,
\begin{align}
    \frac{\mathrm{d}^n}{\mathrm{d}r^n} T_0(r) &= f^{(n)}(r) \,, \\
    \frac{\mathrm{d}^n}{\mathrm{d}r^n} T_{t + 1}(r) &= \sum_{k = 0}^{n + 1} \binom{n + 1}{k} \gamma^{(n+1-l)}(r) T_t^{(k)}(r) \,.
\end{align}
The algorithm for evaluation of the nested ``triangular'' exponential integral of arbitrary dimension \(N\) then begins by calculation of derivatives of \(f_i(r)\) up to some high order dependent on \(M\) and \(P\), followed by recurrent evaluation of terms \(T_t\) and the integrals \(I\) (and their needed derivatives) for subsequently higher dimensions.

Note that, as would be expected from the multi-dimensional nature of the problem, while the asymptotic computational complexity for calculation of a single ``triangular'' integral using this approach is manageable \(O(M^3 P^2)\), the computational complexity of the whole algorithm is asymptotically proportional to \(M!\). This scaling is a result of the \(r_1 < r_2 < ... < r_M\) unique orderings (permutations) are required. The corresponding number of independent triangular integrals to be calculated is proportional to \(2^M\) for all splittings of the regular Coulomb function \(F_l\) into the Coulomb-Hankel functions \(H_l^\pm\), and also up to \(n^M\) (\(n\) being the typical number of channels in an irreducible representation) for evaluation of a multi-dimensional integral for each possible (dipole coupled) sequence of intermediate state outer region channels. Recall that \(M\) refers to the number of photons absorbed in the continuum. Consequently, the problem is computationally tractable only for modest above-threshold ionisation orders.

When the R-matrix radius \(r = a\) is too small for the asymptotic procedures to be valid, it is possible to employ numerical integration between \(r = a\) and some sufficiently large \(r = b\) that is suitable for application of the asymptotic methods. The numerical integration typically needs to be employed for slowly oscillating functions only. Outside of the \((a,b)^M\) hypercube, e.g.\ in \((a,b)^{M-1} \times (b,+\infty)\), it can be combined with the analytic formulas.

Even though they are not necessary---as we have just outlined a general algorithm---for convenience and illustration we include below the explicit forms of the one- and two-dimensional integrals. The one-dimensional case is
\begin{equation}
    I[r^m, \theta]_r = \mathrm{e}^{\mathrm{i}\theta(r)} \gamma(r) \sum_{p = 0}^{P} \mathrm{i}^p r^m
    \frac{\sum_{q = 0}^p a_{mpq} (ur)^{p - q} v^q}{(ur + v)^{2p}} \,,
    \label{eq:I1explicit}
\end{equation}
where \(u = s_1 k_1 + s_2 k_2\) and \(v = Z(s_1/k_1 + s_2/k_2)\). The coefficient \(a_{mpq}\) is defined recursively as
\begin{align}
    a_{m00} &= 1 \,, \\
    a_{mpq} &= (m + 1 - p - q)a_{m,p-1,q} \nonumber \\
            &+ (m + 1 + p - q)a_{m,p-1,q-1} \,,
\end{align}
and \(a_{mpq} = 0\) for invalid indices (\(p < 0\), \(q < 0\) or \(q > p\)). For two dimensions, where additionally \(\phi(r) = s_3\theta_3(r) + s_4\theta_4(r)\), \(\tilde{\gamma}(r) = \mathrm{i}/(\phi'(r) + \theta'(r))\), \(\tilde{u} = s_1 k_1 + s_2 k_2 + s_3 k_3 + s_4 k_4\), \(\tilde{v} = Z(s_1/k_1 + s_2/k_2 + s_3/k_3 + s_4/k_4)\) and \(\tilde{m} = m_1 + m_2\), we have
\begin{align}
    &I\left[r^{m_2} I\left[r^{m_1}, \theta\right], \phi\right]_r = \mathrm{e}^{\mathrm{i}\phi(r) + \mathrm{i}\theta(r)} \tilde{\gamma}(r) \nonumber \\
    & \times \sum_{p = 0}^{P} \mathrm{i}^{p}
    r^{m_1} \frac{\sum_{q = 0}^p a_{m_1pq} (ur)^{p-q} v^q B_{m_1m_2pq}(r)}{(ur+v)^{2p}} \,,
\end{align}
\begin{align}
    &B_{m_1m_2pqs}(r) = \gamma(r) \nonumber \\
    &\times \sum_{s = 0}^{P} \mathrm{i}^s r^{m_2} \frac{\sum_{t,t' = 0}^{s} b_{\tilde{m}pqstt'} (ur)^{s-t} v^{t} (\tilde{u}r)^{s-t'} \tilde{v}^{t'}}{(\tilde{u}r + \tilde{v})^{2s} (ur + v)^{s}} \,,
    \label{eq:I2B}
\end{align}
with the recursive definition of the constant \(b_{\tilde{m}pqstt'}\)
\begin{align}
    b_{\tilde{m}pqstt'}
    &= (\tilde{m} - p - q - s - t - t' + 1) b_{\tilde{m}pq(s-1)tt'} \nonumber \\
    &\ + (\tilde{m} + p - q - t - t' + 2) b_{\tilde{m}pq(s-1)(t-1)t'} \nonumber \\
    &\ + (\tilde{m} - p - q + s - t - t' + 1) b_{\tilde{m}pq(s-1)t(t'-1)} \nonumber \\
    &\ + (\tilde{m} + p - q + 2s - t - t' + 2) b_{\tilde{m}pq(s-1)(t-1)(t'-1)} \,,
\end{align}
with the initial condition \(b_{\tilde{m}pq000} = 1\), and \(b_{\tilde{m}pqstt'} = 0\) whenever \(t < 0\), \(t > s\), \(t' < 0\) or \(t' > s\).

\section{Orientation averaging}
\label{sect:oavg}

The laboratory-frame differential cross section of ionization of a sample of randomly oriented molecules by absorption of \(n\) photons with given laboratory-frame polarizations
\(p_1, \dots, p_n\) ($p_{i}$ is equal to 0 for linear polarization or \(\pm1\) for a circular polarization) is
\begin{equation}
    \frac{\mathrm{d}\sigma}{\mathrm{d}\Omega}
    = 2\pi(2\pi\alpha\omega)^n
    \int\limits_0^{2\pi}
    \int\limits_0^{\pi}
    \int\limits_0^{2\pi} |M(\alpha,\beta,\gamma)|^2
    \frac{\mathrm{d}\alpha \mathrm{d}\beta \mathrm{d}\gamma}{8\pi^2},
    \label{eq:oavg}
\end{equation}
and has the form of a Legendre expansion
\begin{equation}
    \frac{\mathrm{d}\sigma}{\mathrm{d}\Omega}
    = \frac{\sigma_0}{4\pi} \left( 1 + \sum_{L = 1}^{2n} \beta_L P_L(\cos\theta) \right)
\end{equation}
where \(\sigma_0 = b_0\) is the isotropic generalized integral cross section, \(\beta_L = b_L/\sigma_0\) are the dimensionless asymmetry parameters, and the quantities \(b_L\) are
\begin{widetext}
\begin{align}
    b_L &= \sum_{\substack{l_f m_f m_f', \lambda_f \mu_f \\ m_n \dots m_1, \mu_n \dots \mu_1}}
    (-1)^{m_f'} (2L + 1) \sqrt{(2l_f + 1)(2\lambda_f + 1)}
    M_{fi, l_f m_f, m_n \dots m_1} M_{fi, \lambda_f \mu_f, \mu_n \dots \mu_1}^* \nonumber \\
    &\times \sum_{\substack{j_1 \dots j_n, \tilde{j}_1 \dots \tilde{j}_n \\ \nu_1 \dots \nu_n, \nu_1' \dots \nu_n', \tilde{\nu}_1 \dots \tilde{\nu}_n}}
    \left(\begin{matrix} l_f & \lambda_f & L \\  0 & 0 & 0 \end{matrix}\right)
    \left(\begin{matrix} l_f & \lambda_f & L \\ -m_f' & m_f' & 0 \end{matrix}\right)
    \frac{\delta_{j_n \tilde{j}_n} \delta_{\nu_n' \tilde{\nu}_n'} \delta_{\nu_n \tilde{\nu}_n}}{2j_n + 1} \nonumber \\
    &\times (-1)^{m_f+\nu_1+\dots+\nu_n} (2j_1 + 1) \dots (2j_n + 1) \nonumber \\
    &\times
    \left(\begin{matrix} l_f & 1 & j_1 \\ -m_f' & p_1 & -\nu_{1}' \end{matrix}\right)
    \dots
    \left(\begin{matrix} j_{n-1} & 1 & j_n \\ \nu_{n-1}' & p_n & -\nu_n' \end{matrix}\right)
    \left(\begin{matrix} l_f & 1 & j_1 \\ -m_f & m_1 & -\nu_1 \end{matrix}\right)
    \dots
    \left(\begin{matrix} j_{n-1} & 1 & j_{n} \\ \nu_{n-1} & m_n & -\nu_{n} \end{matrix}\right) \nonumber \\
    &\times (-1)^{\mu_f+\tilde{\nu}_1+\dots+\tilde{\nu}_n} (2\tilde{j}_1+1) \dots (2\tilde{j}_n+1) \nonumber \\
    &\times
    \left(\begin{matrix} \lambda_f & 1 & \tilde{j}_{1} \\ -m_f' & p_1 & -\nu_{1}' \end{matrix}\right)
    \dots
    \left(\begin{matrix} \tilde{j}_{n-1} & 1 & \tilde{j}_{n} \\ \nu_{n-1}' & p_n & -\nu_n' \end{matrix}\right)
    \left(\begin{matrix} \lambda_f & 1 & \tilde{j}_{1} \\ -\mu_f & \mu_1 & -\tilde{\nu}_{1} \end{matrix}\right)
    \dots
    \left(\begin{matrix} \tilde{j}_{n-1} & 1 & \tilde{j}_{n} \\ \tilde{\nu}_{n-1} & \mu_n & -\tilde{\nu}_{n} \end{matrix}\right)
     \,.
\end{align}
\end{widetext}
Here, \(M_{fi, l_f m_f, m_n \dots m_1}\) are terms of the partial wave expansion of~\eqref{eq:Mdef}; that is, \(l_f\) and \(m_f\) are the molecular-frame angular quantum numbers of the partial wave, while \(m_1, \dots, m_n\) label the components of the dipole operator in spherical basis in the molecular frame. The expression for \(b_L\) was obtained by: expressing the molecular-frame expansion of \(M_{fi}\) in terms of the laboratory-frame quantum numbers \(m_f'\) and \(p_1, \dots, p_n\),
\begin{align}
    M_{fi} = \sum_{l_f m_f m_f', m_1 \dots m_n} &M_{fi, l_f m_f, m_1 \dots m_n}
        Y_{l_f m_f'}(\hat{\bm{k}}_f) \nonumber \\
    &\times D_{m_f' m_f}^{(l_f)} D_{p_1 m_1}^{(1)*} \dots D_{p_n m_n}^{(1)*} \,;
\end{align}
expressing the product \(Y_{l_f m_f'} Y_{\lambda_f \mu_f'}^*\) that arises in equation~\eqref{eq:oavg} as a series in \(Y_{LM}\); recursively combining the rotational matrices associated with a partial wave with those associated with the dipole operator; and eventually using the orthogonality of the rotational matrices.

\begin{figure}[htbp]
    \centering
    \includegraphics[width=0.5\textwidth]{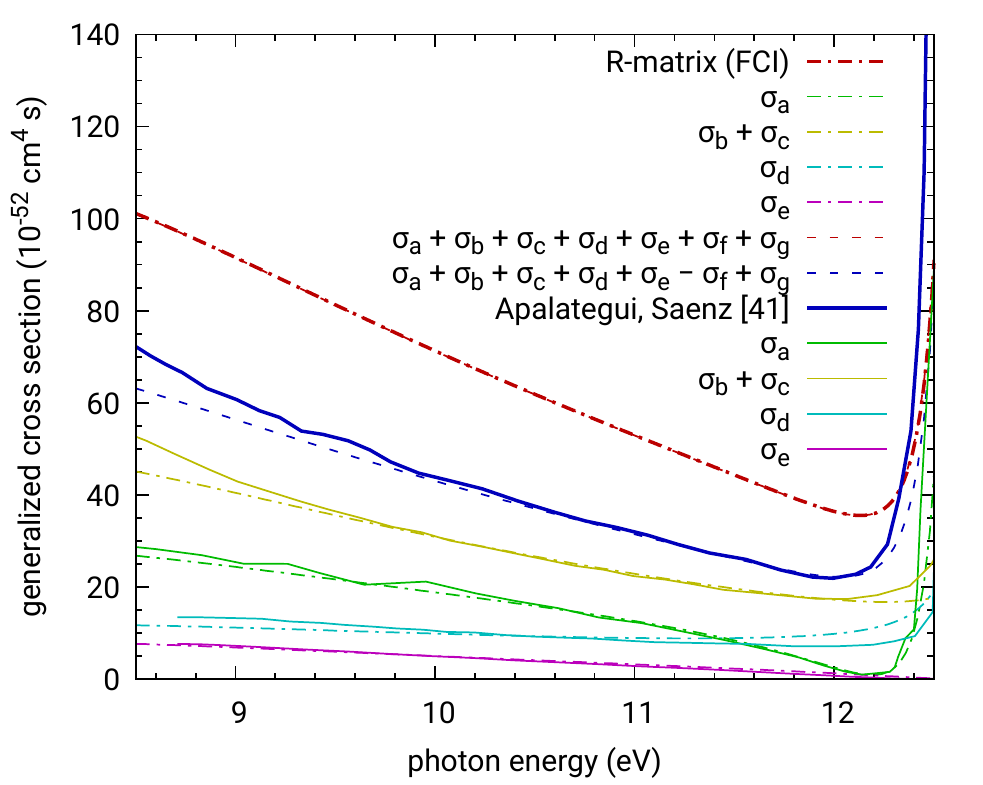}
    \caption{Comparison of components \(\sigma_a, \sigma_b, \dots, \sigma_g\) of the orientation-averaged isotropic total cross section of two-photon ionization of H$_2$ in the FCI model (broken lines) to results of Apalategui and Saenz~\cite{Apalategui_2002} (solid lines).}
    \label{fig:H2components}
\end{figure}

In the main text we mention a discrepancy in Fig.~\ref{fig:H2-unoriented}a between the orientation-averaged isotropic cross section for two-photon ionization of H$_2$ calculated in this work and the results of Apalategui and Saenz~\cite{Apalategui_2002}. The authors of the latter article break down the total cross section into components \(\sigma_a, \sigma_b, \dots, \sigma_g\), some of which they explicitly plot. In Fig.~\ref{fig:H2components} we reproduce those components and demonstrate that if we flip the sign of the interference term \(\sigma_f\) with respect to the definition (7f) in the reference, we obtain their results (as far as our calculated transition matrix elements permit); cf.\ thin broken blue curve to thick solid blue curve in Fig.~\ref{fig:H2components}. With the exact definition of \(\sigma_f\) in the article, however, we obtain our results; the two red curves in Fig.~\ref{fig:H2components} lie on top of each other. Based on this observation, together with the fact that we obtained our results also in Cartesian basis via independent formula~(\ref{eq:nnnn}), which does not treat \(\sigma_f\) separately, we conclude that the authors of the earlier calculation used an incorrect sign of \(\sigma_f\) when evaluating the final orientation averaged cross section and that the agreement between the calculations is then actually much better than Fig.~\ref{fig:H2-unoriented}a suggests. Nevertheless, the formula (7) of their paper appears to be correct.

\end{appendix}

\newpage

\bibliography{bibliography}

\section*{Acknowledgements}

The authors thank Dr. Thomas Meltzer for careful reading of the manuscript and for helpful suggestions. Z.M. and J.B. acknowledge support of the Czech Science Foundation as the project GA CR 20-15548Y and support of the PRIMUS programme of the Charles University as the project PRIMUS/20/SCI/003. Some computational resources were supplied by the project ``e-Infrastruktura CZ'' (e-INFRA LM2018140) provided within the program Projects of Large Research, Development and Innovations Infrastructures. This work was also supported by The Ministry of Education, Youth and Sports from the Large Infrastructures for Research, Experimental Development and Innovations project ``IT4Innovations National Supercomputing Center -- LM2015070''. 

\section*{Author contributions statement}

J.B. independently developed the theory, implemented it into the UKRmol+ code, ran all calculations and wrote the first draft of the manuscript. Z.M. supervised the work and wrote parts of the manuscript.

\section*{Data Availability}

All data generated during this study are included in this published article and its Supplementary Information files.

\section*{Supplementary Information}

The files attached contain all data sets used to produce the figures. A descriptive list of the contents of each file is provided in the \texttt{README.md} file. More information is available from the authors upon a reasonable request.

\section*{Additional information}

The authors declare no competing interests.

\section*{Figure legends}

\begin{itemize}
    \item Figure 1. Illustration of the multi-photon transitions, including above-threshold ionization, calculable by the presented method. From left to right \(N+0\), \(N+1\) and \(N + M\) multi-photon ionization processes. Red arrows mark the \(M\) photons absorbed by the photo-electron ``in continuum''. In the special case of the \([N+M]\) REMPI scheme the first \(N\) photons excite the target to an intermediate bound state and the remaining \(M\) photon absorptions ionize the target without further photon absorptions in the continuum, thus corresponding to our \((N+M)+0\) case.
    \item Figure 2. Left: Generalized cross section of two-photon ionization of the hydrogen atom. The vertical line marks the one-photon ionization threshold. Results are compared to earlier theoretical calculations of Klarsfeld~\cite{Klarsfeld} and Karule~\cite{Karule}. Right: Generalized cross section of two-photon ionization of the helium atom calculated in UKRmol+ with a Gaussian basis set used to represent the states of He$^+$. The solid grey vertical line marks the calculated one-photon single ionization threshold, while the broken yellow vertical line marks the two-photon double ionization threshold at~39.5~eV~\cite{Shakeshaft}. Results are compared to earlier theoretical calculations of Sánchez et al.~\cite{Sanchez}, of Feng and van der Hart~\cite{Feng} and of Shakeshaft~\cite{Shakeshaft}.
    \item Figure 3. One-, two-, three- and four-photon generalized cross section for below- and above-threshold ionisation of the hydrogen molecule by a field polarised parallel with the molecular axis. The panels to the right provide details (from top to bottom) of the two-, three- and four-photon data around the multi-photon ionisation thresholds. The purple, green, blue and yellow vertical chain lines mark the one-, two-, three- and four-photon thresholds, respectively.
    \item Figure 4. Generalized cross section of two-photon ionization of molecular hydrogen calculated using UKRmol+ with the atomic basis set aug-cc-pVTZ and full CI wave function model. Left panel: Field polarized parallel to the molecular axis. Right panel: Field polarized perpendicular to the molecular axis. Below-threshold results are compared to calculations of Apalategui and Saenz~\cite{Apalategui_2002} and Morales et al.~\cite{Morales09} The calculated one-photon ionization thresholds as well as the calculated positions of core-excited resonances allowed by symmetry are marked by grey vertical lines (solid and chain, respectively). The yellow broken line marks the calculated vertical non-sequential two-photon double ionization threshold at~25.7~eV.
    \item Figure 5. Laboratory-frame photoelectron angular distribution parameters for two-photon ionization of H$_2$. Left panels: With the same full CI model as in Fig.~\ref{fig:H2-ATZ-FCI-oriented}. Right panels: With the SE model as in Fig.~\ref{fig:H2-multiphot}. Top panels: Averaged cross sections over all relative orientations of the molecule and the field polarization. Middle and bottom panels: Dimensionless asymmetry parameters. The circle marks the theoretical result of Ritchie and McGuire~\cite{RitchieMcGuire}, chain curve the results of Apalategui and Saenz~\cite{Apalategui_2002} and the dashed curve the results of Demekhin et al.~\cite{Demekhin} The disagreement between our results and those of Apalategui and Saenz~\cite{Apalategui_2002}, possibly arising due to a typo in their codes, is discussed in Appendix~\ref{sect:oavg}.
    \item Figure 6. Laboratory-frame photoelectron angular distribution parameters for below- and above-threshold two-photon ionization of CO$_2$ into its first four ionic states. The solid grey vertical line marks the first one-photon ionization threshold at 13.78~eV, while the dashed lines mark further calculated one-photon thresholds for states A, B, C and D, indicating narrow energy windows with inaccurate results.
    \item Figure 7. Summed two-photon isotropic partial ionization cross sections of CO$_2$ including the final states X, A, B, and C. The resonance structure in the below-threshold two-photon ionization is compared to below-threshold one-photon absorption measurement of Chan et al.~\cite{ChunCO2} The grey vertical line marks the one-photon ionization threshold. The first two resonances corresponding to dipole-allowed excitation of the neutral molecule to the singlet excited states \(^1\Sigma_u\) and \(^1\Pi_u\) are labeled in the plot.
    \item Figure 8. Real and imaginary part of the stationary intermediate-state wave function for above-threshold ionization of the hydrogen atom after absorption of one photon in the continuum for a specific energy. The vertical line as well as the labels ``(inner)'' and ``(outer)'' mark the division into R-matrix regions.
    \item Figure 9. Comparison of components \(\sigma_a, \sigma_b, \dots, \sigma_g\) of the orientation-averaged isotropic total cross section of two-photon ionization of H$_2$ in the FCI model (broken lines) to results of Apalategui and Saenz~\cite{Apalategui_2002} (solid lines).
\end{itemize}

\end{document}